\documentclass[preprint]{aastex}

\pdfoutput=1

\usepackage{ifthen,graphicx}

\newcommand{\mysize}{\normalsize}

\newcommand{\mytoday}{%
        \number\day~\ifcase\month\or January\or February\or 
        March\or April\or May\or June\or July\or August\or 
        September\or October\or November\or December\fi~\number\year}

\newcommand{\etal}{{\it et al.,\ }}
\newcommand{\etaldot}{{\it et al.\ }}
\newcommand{\msc}[1]{{\mbox{\scriptsize{#1}}}}
\newcommand{\msct}[1]{{\mbox{\tiny{#1}}}}

\newcommand{\mar}[1]{\marginpar{\vspace{15pt}\textsf{\footnotesize #1}}}

\renewcommand{\L}[1]{L\ensuremath{_\msc{#1}}}

\newcommand{\fun}[1]{\ensuremath{f^\msct{UN}_\msct{#1}}}
\newcommand{\fnt}[1]{\ensuremath{f_\msct{#1}}}

\newcounter{mypage}

\newcounter{myfigs}

\newcommand{\mylabel}[1]{
        \setcounter{mypage}{\value{page}}
        \addtocounter{mypage}{-1}
        \refstepcounter{mypage}
        \label{#1}}

\begin{document}

\renewcommand{\L}[1]{L\ensuremath{_\msc{#1}}}

\title{\textmd{\small \textsl{Icarus} manuscript \#11785, submitted 8 February 2011}\\[1cm]
Transformation of Trojans into Quasi-Satellites During  \\
Planetary Migration and Their Subsequent  \\
Close-Encounters with the Host Planet \\[1cm]} 

\author{
Stephen J. Kortenkamp and Emily C. S. Joseph \\
Planetary Science Institute, Tucson, AZ\\
kortenka@psi.edu}

\normalsize

\section*{\centering {\normalsize ABSTRACT}}

\noindent
We use numerical integrations to investigate the dynamical evolution of resonant Trojan and quasi-satellite companions
during the late stages of migration of the giant planets Jupiter,
Saturn, Uranus, and Neptune.  Our migration simulations begin with Jupiter and Saturn on orbits already well separated from their mutual 2:1 mean-motion resonance.  Neptune and Uranus are decoupled from each other and have orbital eccentricities damped to near their 
current values.   
From this point we adopt a planet migration
model in which the migration speed decreases exponentially with a
characteristic timescale $\tau$ (the e-folding time).  We perform a
series of numerical simulations, each involving the migrating giant
planets plus test particle Trojans and quasi-satellites.  
We find that the libration frequencies of Trojans are similar to those of quasi-satellites.  This similarity enables a dynamical exchange of objects back and forth between the Trojan and quasi-satellite resonances during planetary migration.  This exchange is facilitated by secondary resonances that arise whenever there are multiple migrating planets.  For example, secondary resonances may occur when the circulation frequencies, $f$, of critical arguments for the Uranus-Neptune 2:1 mean-motion near-resonance are commensurate with harmonics of the libration frequency of the critical argument for the Trojan and quasi-satellite
1:1 mean-motion resonance (e.g., \fun{2:1}\,=\,2\,\fnt{1:1}).  
Furthermore, under the influence of these secondary resonances quasi-satellites can have their libration amplitudes enlarged until they undergo a close-encounter with their host planet and escape from the resonance.  High-resolution simulations of this escape process reveal that 
$\approx$80\% of Jovian quasi-satellites experience one or more close-encounters within Jupiter's Hill radius (R$_\msct{H}$) as they are forced out of the quasi-satellite resonance.  As many as $\approx$20\% come within R$_\msct{H}$/4 and $\approx$2.5\% come within R$_\msct{H}$/10.  Close-encounters of escaping quasi-satellites occur near or even below the 2-body escape velocity from the host planet.  
Finally, the exchange and escape of Trojans and quasi-satellites continues to as late as 6-9$\tau$ in some simulations.  By this time the dynamical evolution of the planets is strongly dominated by distant gravitational perturbations between the planets rather than the migration force.   This suggests that exchange and escape of Trojans and quasi-satellites may be a contemporary process associated with the present-day near-resonant configuration of some of the giant planets in our solar system.

\newpage

\vspace{-0.5cm}
\section*{\normalsize 1 \  \ INTRODUCTION}
\vspace{-0.2cm}

The 1:1 co-orbital resonance is usually discussed in the context of Trojan or horseshoe objects and their association with Lagrangian equilibrium points.  Examples include the many Trojan asteroids locked in resonance with Jupiter, or the Saturnian satellites Janus and Epimetheus, which are on horseshoe
orbits with respect to each other.   These objects are precluded from having relatively close encounters with the secondary object hosting the resonance (the planet or the other horseshoe object).  In fact, this is generally the case in nearly all resonant configurations.  However,  there is another class of 1:1 resonant objects in which the opposite is true.  Known as quasi-satellites, these objects always remain very near to their host planets.  

The quasi-satellite resonance is somewhat unusual and is not discussed in standard dynamics books (e.g., Murray \& Dermott,
1999).  It was first described a century ago (Brown, 1911; Jackson, 1913) and has been steadily gaining attention in other work 
(e.g., H\'enon, 1969; Peterson, 1976; Namouni, 1999; Wiegert \etal 2000; Nesvorn\`y \etal 2002; Connors \etal 2002, 2004; Kortenkamp, 2005; Wajer, 2009, 2010).  Quasi-satellites recently leapt from the realm of mere theoretical curiosities to the real world  with the discovery of near-Earth objects 2002\,AA29  (Connors \etal 2002), 2003 YN107 (Connors \etal 2004), 2004 GU9 (Weigert, 2005), and 2006 FV35 (Wajer, 2010). Both 2004 GU9 and 2006 FV35 are currently natural quasi-satellites of Earth and are predicted to remain so forÊ about 1000 and 800 years, respectively (Wajer, 2010). Asteroid 2003 YN107 occupied the quasi-satellite resonance from 1997-2006, but is now a horseshoe object (Connors \etal 2004). Near-earth object 2002 AA29 is also presently on a horseshoe orbit, but numerical simulations indicate that it was a quasi-satellite object approximately 500 years ago, and will re-enter the resonance around the year 2600 (Connors \etal 2002; Wajer, 2009).

\mar{Fig.~\ref{FIG:ref_frames}}

Figure~\ref{FIG:ref_frames} demonstrates the orbital characteristics of quasi-satellites in two different reference frames, one rotating with the same average motion of the planet (Fig.~\ref{FIG:ref_frames}{A}) and the other a non-rotating planet-centered frame (Fig.~\ref{FIG:ref_frames}{B}).  The dotted path in Figure~\ref{FIG:ref_frames}{A} indicates the motion of a quasi-satellite with orbital eccentricity larger than the planet's eccentricity.  In Figure~\ref{FIG:ref_frames}{B} and throughout this paper we use the Hill radius, R$_\msc{H}$, to indicate the distance from a planet where solar gravity and the planet's gravity are approximately equal.  This relation is expressed as R$_\msc{H} = a_\msc{pln} \left[{m_\msc{pln}}/({3 M_\odot})\right]^{1/3}$, where $m_\msc{pln}$ and $a_\msc{pln}$ are the planet's mass and semi-major axis, respectively, and $M_\odot$ is the sun's mass.  Inside the Hill radius an object's motion is dominated by the planet, while beyond R$_\msc{H}$ its motion is dominated by the sun.   All true satellites, regular and irregular, orbit deep inside their planet's Hill radius.  Earth's moon, for example, orbits at about one-quarter of Earth's Hill radius.

The name quasi-satellite is fitting when one considers the behavior of these objects in the rotating reference frame of Figure~\ref{FIG:ref_frames}{A}.  In this frame the quasi-satellites 
appear to be orbiting both the sun and the planet with the same period.  That is, at one point they are between the planet and the sun, while half an orbit later they are in the anti-sun direction.   In an inertial reference frame (not shown) the differences in orbital eccentricities and longitudes of pericenter of the quasi-satellites and the planet result in the planet leading the quasi-satellite for half an orbit and then being overtaken and lagging behind it for the other half-orbit.  As a consequence, as seen from the host planet (Fig.~\ref{FIG:ref_frames}{B}) the quasi-satellites do not appear
to be in orbit around the planet at all.  Rather, they remain
in the same quadrant of the sky and, in this example, are never much more than about 10 R$_\msc{H}$ away from the planet.  To an observer stationed on or near Earth a terrestrial quasi-satellite would be observable in only one specific area of the sky throughout the year. 

In this paper we demonstrate that there can be a dynamical exchange of objects between the Trojan and quasi-satellite resonances.  The exchange occurs during the late stages of planetary migration, as the giant planets are approaching their current orbital configuration.  Section 2 below describes our numerical methods and initial conditions.  In Section 3 we describe details of the Trojan to quasi-satellite exchange process.  Section 4 describes the continuing effects of planetary migration on quasi-satellites after they are in the resonance.  We conclude in Section 5 by discussing some implications of this work for other populations of small solar system bodies.

\vspace{-0.5cm}
\section*{{\normalsize 2  \  \ METHODS}} 
\vspace{-0.2cm}

In a previous paper (Kortenkamp \etal 2004) we investigated the survivability of Trojan-type companions of Neptune during primordial radial migration of the giant planets Jupiter, Saturn, Uranus, and Neptune.  A shortcoming of this earlier work is that the Trojans were removed from the simulations when either; (1) they suffered a close-encounter with Neptune, or (2) their mean longitudes differed from that of Neptune by more than 180$^\circ\!\!$, meaning they had left the Trojan region.  The first of these reasons was a programmatic necessity, as the integrator being used could not accurately handle close-encounters.  
The second was a crude attempt to decrease the CPU requirements of the simulations.  
In addition, each simulation involved only $\sim$1000 Trojans.  Because of these limitations, transitions between the Trojan and quasi-satellite resonances would not have been readily observed.  In this paper we overcome these problems by using an integrator designed to handle close-encounters, and following an order of magnitude more particles for the entire duration of each simulation unless they were cleared from the planetary system.

\subsection*{\normalsize 2.1  \  \  Numerical Model}

Our simulations were performed using the {\em Mercury}\ hybrid $N$-body code of
Chambers (1999).  In {\em Mercury}, Chambers has combined a Wisdom-Holman (1991) symplectic
integration algorithm (Levison \& Duncan, 1994) with a Bulirsch-Stoer
(Stoer \& Bulirsch, 1980) integrator for handling close-encounters.  The {\em Mercury} package also allows for a user-defined force to be added to both the symplectic component and the general Bulirsch-Stoer close-encounter component.  For the modeling described in this paper a smooth drag force was added to cause the radial migration of the planets by inducing a time variation of the planets'
semi-major axes, $a$.  Following Malhotra (1995) a timescale $\tau$ was used to characterize
the migration of semi-major axis $a$ as a function of time $t$, where $a(t) = a(0) + \Delta a [1 - \exp(-t/\tau)]$ and
$\Delta a$ is the desired amount of total migration at time $t = \infty$.  Our nominal model used $\tau=10^6$ years while some simulations used a slower migration, with $\tau=10^7$ years.

The mutual gravitational perturbations of the planets were included in the simulations self-consistently even as their orbital spacing was expanding.  The numerical simulation code is similar to that used in our earlier work (Kortenkamp \& Wetherill, 2000; Kortenkamp \etal 2001; Kortenkamp, 2005).  A small number of simulations were simultaneously carried out with the same initial conditions but using the alternative code used previously in Kortenkamp \etaldot (2004).  The legacy of this alternative integrator dates to Malhotra (1995) and is distinct from {\em Mercury}.  There was general agreement between the outcome of the two codes when close-encounters with planets were not involved.

Determining an appropriate time step to use for symplectic integrations of quasi-satellites is non-trivial.  We conducted a limited number of idealized simulations that included only a single non-migrating giant planet and an initial population of quasi-satellites.  Time steps in the range of 1/6$^\msc{th}$ to 1/10$^\msc{th}$ the planet's orbital period resulted in rapid artificial escape of all pre-existing quasi-satellites.  For time steps $\leq$1/20$^\msc{th}$ the host planet's orbital period all initial quasi-satellites remained stable for $\sim$10$^7$ years.  For the work described here this time step issue was critically important for simulations of Jovian quasi-satellites, less so for those of Neptune.  In the latter case the presence of the inner giant planets in the simulations becomes the controlling factor in choosing an appropriate time step.  For all simulations in which Jupiter was the host planet we used a time step of 6 months, while for those with Neptune as the host we used a time step of 1 year.

\subsection*{\normalsize 2.2  \  \  Initial Conditions of the Giant Planets}  

In the Nice model for the very early evolution of the solar system (Morbidelli \etal 2005; Gomes \etal 2005; Tsiganis \etal 2005; Levison \etal 2008) it is proposed that the four giant planets Jupiter, Saturn, Uranus, and Neptune may have formed in a compact configuration (all within about 15 AU of the sun).  After their formation, they underwent significant orbital migration caused by the scattering of remnant planetesimals.  During this process the planets exchanged angular momentum with the planetesimals and thus their orbits migrated radially inward or outward (for a detailed discussion of the planetesimal migration process see Hahn \& Malhotra, 1999). One primary difference between the Nice model and alternative planetary migration models is that the Nice authors propose that Saturn formed closer to the sun than Jupiter's outer 2:1 mean-motion resonance.  Thus, during the primordial migration phase the outwardly-migrating Saturn would have crossed this major resonance to reach its current configuration relative to Jupiter.  Crossing of their mutual 2:1 resonance by Jupiter and Saturn would have dislodged any primordial population of Trojans.  The capture of Jupiter's current population of Trojan asteroids may then have occurred as Jupiter and Saturn diverged from the 2:1, and associated secondary resonances swept planetesimals into the 1:1 resonance region (Morbidelli \etal 2005).  This process, called chaotic capture, may also have allowed Neptune to acquire its Trojan swarms via secondary resonances associated with Uranus (Nesvorn\`y \& Vokrouhlick\`y, 2009) as these two planets migrated.  Regardless of the validity of the initial conditions in the Nice or any other model, it is likely that as the giant planets were approaching their final orbits during the late stages of planetary migration both Jupiter and Neptune would have hosted significant populations of Trojan objects with orbital distributions not too unlike the present populations.

\mar{Tab.~\ref{tab:model_parameters}}

Our migration simulations began with Jupiter and Saturn on orbits already separated from their mutual 2:1 mean-motion resonance.  We also included Uranus and Neptune on orbits decoupled from each other and with orbital eccentricities damped to near their current values.  Thus, we were modeling the very late stages of giant planet migration.  Table~\ref{tab:model_parameters} lists the masses and initial configuration of these planets for our simulations.  However, we note that our specific initial orbital distribution is not critical to the general outcome.  As long as the planets migrated towards their current configuration, the secondary resonances discussed below will affect the Trojan and quasi-satellite objects. 

\mar{Fig.~\ref{FIG:planetary_orbits}}

Figure~\ref{FIG:planetary_orbits} shows the evolution of the semi-major axes of
the four giant planets as a function of time expressed in units of the
characteristic migration timescale $\tau$ up to a time of $8\tau$.
With this exponential migration model, radial migration of all planets
is 99.75\% complete after a time of $6\tau$.  By this time orbital
evolution of the planets is dominated
by mutual planetary gravitational perturbations rather than the effects of the migration force.  The top panel of Figure~\ref{FIG:planetary_orbits} also shows the location of the outer 2:1 mean-motion resonance with Uranus.  As migration proceeds, Uranus and Neptune asymptotically approach this 2:1 resonant configuration but never quite reach it.  The bottom panel of Figure~\ref{FIG:planetary_orbits} similarly shows Jupiter approaching the location of the inner 2:5 mean-motion resonance with Saturn.

\subsection*{\normalsize 2.3  \  \  Initial Conditions of Trojans and Quasi-satellites}  

Two different types of simulations were performed.  One type involved initial populations of Trojan objects associated either with Jupiter or Neptune.  The other type involved initial populations of quasi-satellites, again either Jovian or Neptunian.   In all cases the Trojans and quasi-satellites were modeled as massless test particles that were not subject to the migration-inducing drag force acting on the planets. The initial orbital distributions of these test particles were obtained as follows.

\mar{Fig.~\ref{FIG:initial_trojans}}

For initial conditions of the Jupiter and Neptune Trojans we started with the
population of Jupiter's real Trojan companions,
a total of 2262 asteroids as of 28 September, 2007\footnote{The orbital elements
of Jupiter's Trojans were downloaded from the Minor Planet Center database at cfa-www.harvard.edu/iau/lists/JupiterTrojans.html.}$\!\!$.  The distribution in orbital elements of these real Trojans  were then used to randomly generate $10^4$  Trojan particles evenly divided between the leading \L{4} and trailing \L{5} regions (see appendix for a note on this naming convention).   The initial Trojan population resulting from this process is shown in Figure~\ref{FIG:initial_trojans}.  The distributions in the differences in semi-major axis ($a$), argument of pericenter ($\omega$), and longitude of ascending node ($\Omega$) of these Trojan test particles with respect to the real Jupiter were then used to shift the test particle population to the initial orbits of the simulated Jupiter or Neptune. 

For those simulations that began with an initial population of quasi-satellites the test particle orbital eccentricities were randomly distributed in the range $e = 0.15-0.35$ for Jovian quasi-satellites and in the range $e = 0.05-0.15$ for Neptunian quasi-satellites.  For both populations orbital inclinations were randomly distributed between 0 and 25$^\circ\!\!$.  These values are generally consistent with the regions of stability found in simulations of quasi-satellite evolution using the current giant planet orbits (Wiegert \etal 2000).  For all quasi-satellite test particles $\omega$ and $\Omega$ were both randomly assigned between 0 and 360$^\circ\!\!$, with the longitude of pericenter then being $\tilde{\omega} = \omega + \Omega$.  Quasi-satellite semi-major axes and mean longitudes ($\lambda$) were set equal to the initial simulated $a$ and $\lambda$ of either Jupiter or Neptune.  Finally, the mean anomaly ($M$) was found from $M = \lambda - \tilde{\omega}$.

\mar{Fig.~\ref{FIG:qs_hires}}

The condition for identifying an object as
trapped in a 1:1 resonance with a planet is libration of the critical argument $\phi_\msc{1:1} = \lambda - \lambda_\msct{pln}$.  For non-resonant objects,  $\phi_\msc{1:1}$ will circulate through all angles from 0 to 360$^\circ\!\!$.  Objects trapped as Trojans have $\phi_\msc{1:1}$ librating back and forth around either $+60^\circ$ or $-60^\circ\!\!$,\,\ while $\phi_\msc{1:1}$ for horseshoe objects librates with a very large amplitude around 180$^\circ\!\!$.\,\   Quasi-satellites have $\phi_\msc{1:1}$ librating around 0$^\circ\!\!$.\,\  Figure~\ref{FIG:qs_hires} demonstrates this libration for a test particle trapped in Neptune's quasi-satellite resonance over a $10^4$ year period during one of our simulations.  High-frequency oscillations in $\phi_\msct{1:1}$ occur at about 6 cycles per $10^3$ years (kyr$^{-1}$; see spectrum in bottom panel of Fig.~\ref{FIG:qs_hires}).  This high frequency libration corresponds to the quasi-satellite's orbital period around the sun.  It is also the frequency at which the quasi-satellite traces out the double-lobed looping path shown in Figure~\ref{FIG:ref_frames}{B}\@.  The quasi-satellite's libration with respect to the planet also has components at much lower frequency, corresponding to a libration period in the range of 2-5$\times10^3$ years.  

\vspace{-0.5cm}
\section*{{\normalsize 3  \  \  TURNING TROJANS INTO QUASI-SATELLITES}}
\vspace{-0.2cm}

During the late stages of planetary migration Neptune and Uranus were approaching their mutual 2:1 mean-motion resonance (see Fig.~\ref{FIG:planetary_orbits}).  One critical argument for this resonance is $\phi^\msct{UN}_\msct{2:1}=2\lambda_\msct{Nep} - \lambda_\msct{Ura} - \tilde{\omega}_\msct{Ura}$.  Because Uranus and Neptune are not actually in the 2:1 resonance $\phi^\msct{UN}_\msct{2:1}$ circulates through all angles, with a circulation frequency \fun{2:1}.  Secondary resonances with Trojans occur when this circulation frequency becomes commensurate with harmonics of the libration frequency, \fnt{1:1} (e.g., \fun{2:1}\,=\,2\,\fnt{1:1}).  Kortenkamp \etaldot (2004) showed that during planetary migration the dynamical evolution of Trojans is exclusively controlled by these types of secondary resonances associated with the host planet and the next nearest giant planet (in semi-major axis).  Because of these types of secondary resonances, nearly all Trojans of Neptune escape from the 1:1 resonance region during planetary migration and obtain free heliocentric orbits.  Results from our new simulations with the {\em Mercury} integrator confirm this finding.  However, note that the lower frequency component in the libration of Neptune's quasi-satellites shown in Figure~\ref{FIG:qs_hires} is broadly commensurate with the circulation frequency \fun{2:1}.   This suggests that the secondary resonances can also provide a dynamical pathway between Trojans and quasi-satellites, which is shown by our new simulations.

\mar{Fig.~\ref{FIG:new_ejection}}

Figure~\ref{FIG:new_ejection} shows an example from a $\tau = 10^6$ year
simulation where a Neptune Trojan particle was influenced by several secondary resonances between \fun{2:1} and \fnt{1:1}.  A fast Fourier
transform (FFT) was used to obtain power spectra of
$\phi_\msct{1:1}$ (middle panel) and
$\phi^\msct{UN}_\msct{2:1}$ (bottom panel).  Spectra were taken roughly every
$0.15\,\tau$ from about 1.1 to $3.1\tau$.  Each FFT used 2048
points sampled every 50 years, giving an FFT interval of about
$0.1\tau$.  In the first interval, starting at 1.1$\tau$, \fun{2:1}
is higher than the 4th harmonic of \fnt{1:1}.  As Uranus and Neptune
converge upon their 2:1 resonance \fun{2:1} slows and passes the
$4^\msc{th}$ harmonic of \fnt{1:1} at about 1.3$\tau$.  The libration
amplitude of the Trojan particle experiences a perturbation at this
time because of the 1:4 secondary resonance (\fun{2:1} = 4\fnt{1:1}).  Passage of \fun{2:1} by the $3^\msc{rd}$ harmonic of \fnt{1:1}
at about 1.9$\tau$ results in a more significant perturbation to the libration
amplitude.  At about $2.6\tau$ the 2:5 secondary resonance is reached, 
where 2$\fun{2:1} = 5\fnt{1:1}$.  
As a consequence the Trojan is quickly forced to larger libration amplitude.
Increasing libration amplitude results in a  slower libration
frequency.  Thus, the $5^\msc{th}$ harmonic of the slowing \fnt{1:1}
keeps pace with the $2^\msc{nd}$ harmonic of the slowing \fun{2:1}.  In other words, the Trojan is trapped in
the 2:5 secondary resonance.  The secondary resonance forces the particle briefly into the horseshoe regime at about 2.8$\tau$ and then into the quasi-satellite region at about 2.85$\tau$.  At this point the libration decouples from \fun{2:1} and the particle remains a relatively stable quasi-satellite for the next $5\times10^5$ years.

A similar transformation of Trojans into quasi-satellites occurred in the simulations with an initial population of Jupiter Trojans.  In these simulations, secondary resonances associated with Jupiter approaching Saturn's inner 2:5 mean-motion resonance were likely responsible, although effects of other Saturnian resonances, such as the 3:7, could also be important.  

\mar{Fig.~\ref{FIG:qs_trapping_exp}}

Figure~\ref{FIG:qs_trapping_exp} demonstrates the diversity of pathways that Trojans follow to enter the quasi-satellite region.  Many examples were found of quasi-satellites originating from the leading \L{4} or trailing \L{5} regions, as well as from horseshoe orbits.  Occasionally an object escaped from the Trojan region onto a free heliocentric orbit but retained a semi-major axis very close to the planet.   These objects then represented targets of opportunity as secondary resonances with the now circulating $\phi_\msct{1:1}$ were able to trap the objects as quasi-satellites (see panel E in Fig.~\ref{FIG:qs_trapping_exp}).  This is reminiscent of the chaotic capture process for Trojans, where free heliocentric particles in the right place at the right time were susceptible to secondary resonances forcing them into the Trojan regions of a migrating Jupiter (Morbidelli \etal 2005) or Neptune (Nesvorn\`y \& Vokrouhlick\`y, 2009).  The example in panel F demonstrates a fundamental characteristic of secondary resonances --- they can facilitate transitions in either direction between the quasi-satellite and Trojan regions.

With $\tau = 10^6$ years, during the first $3\tau$ of the simulation about 0.2\% of the escaping Jupiter Trojan particles spent some length of time as quasi-satellites after leaving the Trojan region (5 particles out of $\approx$2500 Trojans that escaped).  During the same interval of $3\tau$ Neptune lost less than half as many Trojan particles but about 7\% of those became temporary quasi-satellites (70 particles out of $\approx$1000 Trojans that escaped).  With $\tau = 10^7$ years Jupiter's Trojan losses were more dramatic, with 25\% of the initial population escaping in just 0.2$\tau$, but the rate of quasi-satellite trapping was comparable to the run with $\tau=10^6$ years (4 quasi-satellites from the $\approx$2500 escaping Trojans).  While Neptune's Trojan losses were also elevated in simulations with the slower $\tau = 10^7$ years migration rate, the incidence of quasi-satellite trapping of the escaping Trojans was minimal, with only 3 particles becoming trapped out of the many thousands of escaping Trojans.  Kortenkamp \etaldot (2004) showed that slower migration made it more likely that Trojans would be swept up by the secondary resonances that drive them out of the 1:1 resonance.  Our new results suggest that under slow migration the particles are not able to efficiently decouple from the secondary resonances when they are in the quasi-satellite region.  
 
\vspace{-0.5cm}
\section*{{\normalsize 4  \  \  EVOLUTION OF QUASI-SATELLITES DURING MIGRATION}}
\vspace{-0.2cm}

The results of the previous section indicate that the giant planets likely hosted populations of quasi-satellites while they were migrating.  The fraction of these populations that were ex-Trojans versus captured from free heliocentric orbit is unknown and remains the subject of future work.  Nevertheless, the orbital evolution of quasi-satellites during planetary migration can be studied by generating a pre-existing population as described in Section 2.3 above. 

\subsection*{\normalsize 4.1  \  \  Effects of Secondary Resonances}

\mar{Fig.~\ref{FIG:1_4_planets_nep}}

Objects trapped in the quasi-satellite resonance continue to be affected by secondary resonances, with their libration amplitudes increasing or decreasing in a manner similar to the Trojan objects.  This is because the libration frequencies of trapped quasi-satellites (Fig.~\ref{FIG:qs_hires}) are still broadly commensurate with the circulation frequencies associated with a giant planet relatively near the host planet (in semi-major axis).   In fact, it is {\em only} the nearest giant planet that affects the quasi-satellite, regardless of the different masses of the planets in the system.  To illustrate this we performed additional simulations of quasi-satellite evolution with one or more giant planets removed.  For example, Figure~\ref{FIG:1_4_planets_nep} shows that without Uranus in the simulation (top two rows), Neptune's quasi-satellites remain dynamically stable throughout the entire migration process.  Their libration amplitudes are nearly constant (left column) and they maintain essentially fixed minimum encounter distances (middle column) and relative velocities with respect to Neptune (right column).  The only significant changes in the orbital characteristics shown in the top two rows of Figure~\ref{FIG:1_4_planets_nep} result from the smooth change in Neptune's Hill radius as the planet migrates outward.

Despite the fact that the magnitude of gravitational perturbations from Uranus acting on Neptune's quasi-satellites are dwarfed by perturbations from Jupiter, the dominant role of Uranus in this problem is not entirely unexpected.  Consider the idealized case of the high eccentricity quasi-satellites of Neptune ($e \approx 0.15$).
 From the perspective of one of these quasi-satellites Uranus is within 3.95\,AU at inferior conjunction
and is as distant as 50\,AU at superior conjunction.  Neptune, on the other
hand, is always between 4.39 and 8.75\,AU away from the quasi-satellite.  Accounting for the
difference in planetary mass, the gravitational force from Uranus
acting on Neptune's quasi-satellites can vary from about 1.04 times stronger than the
force from Neptune itself to about 160 times weaker.  These Uranian perturbations will repeat roughly with the synodic
period of Uranus and Neptune.  For a quasi-satellite with non-zero
libration amplitude, its position will vary somewhat at each
conjunction with Uranus.  Thus, sometimes the Uranian perturbations will accelerate and other times
decelerate the particle's motion with respect to Neptune.  The net
effect over long periods would not dramatically alter the quasi-satellite's resonant
configuration with Neptune.  However, if the orbital orientation of Uranus, Neptune, and the quasi-satellite
becomes periodic in time (a secondary resonance)
then the Uranian perturbations become synchronized and drive
the quasi-satellite to either a smaller (more stable) libration amplitude or a larger (less stable) amplitude.  These two effects can be seen in the bottom row of Figure~\ref{FIG:1_4_planets_nep}, where at specific times the influence of Uranus sometimes sharply changes the quasi-satellite's libration amplitude.  These times correspond with the passage of secondary resonances.

For the same idealized Neptune quasi-satellite described above note that Saturnian and Jovian perturbations can be much stronger
than those from Uranus.  Under the initial conditions in
Table~\ref{tab:model_parameters} gravitational perturbations from Jupiter reach about 1.15 times Neptunian at inferior conjunction and only drop to about 0.25 times at superior conjunction, while from Saturn they would respectively be 0.69 and 0.62 times Neptunian.
However, the evolution of Neptune quasi-satellites in simulations that did not include
Uranus (top two rows of Fig.~\ref{FIG:1_4_planets_nep}) demonstrates that the direct Saturnian and
Jovian perturbations have essentially no effect on the stability of Neptune quasi-satellites
during planetary migration.  Jupiter and Saturn have very short synodic
periods with respect to Neptune (about 13 and 35 years, respectively) while Neptune's quasi-satellites
librate with a period $\sim$10$^4$ years.  So it is the frequency of the perturbations from the other planets that is most important, not the magnitude.

The results shown in Figure~\ref{FIG:1_4_planets_nep} and the simple schematic described above do not preclude destabilizing roles for Jupiter and Saturn acting indirectly through perturbations on Uranus, as Michtchenko \etaldot (2001) argued may have been the case for primordial escape of some Neptune Trojans during planetary migration.  Additional simulations similar to those shown in Figure~\ref{FIG:1_4_planets_nep} revealed that the evolution of Jovian quasi-satellites is only significantly affected by secondary resonances associated with Saturn.  

\subsection*{\normalsize 4.2  \  \  Close-Encounters with Planets by Escaping Quasi-Satellites}

In previous work involving quasi-satellite dynamics in the early solar nebula (Kortenkamp, 2005) we showed that slow dissipation of a heliocentric gas disk forced quasi-satellites to evolve closer and closer to the host planet.  Eventually the continuing effects of solar nebula gas drag acting on the quasi-satellites caused them to undergo close-encounters with the planet deep within its Hill radius.  Often these close-encounters occurred at velocities significantly below the 2-body escape velocity.  By this process some quasi-satellites were able to be captured by the planet as true satellites.  For the initial conditions used in the current paper the solar nebula is presumed to have long since dissipated.  However, we were curious to determine if gradual migration of the planets' orbits could lead to similar encounters with quasi-satellites.  In this case, the escape process is totally different.  Instead of gas drag decreasing quasi-satellite libration amplitudes until they encounter the planet, here, the continuing influence of secondary resonances increases the libration amplitudes.  Nevertheless, notice that as the quasi-satellite shown in the bottom row of Figure~\ref{FIG:1_4_planets_nep} escapes (at about 7.8$\tau$) it undergoes a close-encounter with Neptune within the planet's Hill radius, R$_\msct{H}$.  Furthermore, the relative velocity at closest approach is quite low, only about 25\% above the escape velocity from R$_\msct{H}$.

\mar{Fig.~\ref{FIG:5tau_nep}}

We conducted high-resolution simulations with the initial populations of quasi-satellites and confirmed that many quasi-satellites undergo deep low-velocity encounters with the host planet as they escape the resonance or very shortly thereafter.  For these simulations we transitioned to the Bulirsch-Stoer close-encounter integrator when the quasi-satellites approached within 10 R$_\msct{H}$ of the host planet (the nominal switchover distance was 3 R$_\msct{H}$).  For each close-encounter we stored the fine details of the position and velocity of the quasi-satellite relative to the planet at each sub time step within the Bulirsch-Stoer algorithm.  Then, using the coarse details of the orbital elements, we determined when the particle left the quasi-satellite resonance and examined the details of the high-resolution close-encounter data during this period.  Figure~\ref{FIG:5tau_nep} shows close encounter details for five examples from a simulation with Neptune quasi-satellites where the migration timescale was $\tau = 10^7$ years.  Encounters within R$_\msct{H}$ are indicated by one of the three different types of grey circles in the plot (see legend at top of middle row).  The corresponding relative velocities at these deep encounters are indicated in the plots of the right column.  Many of the escaping quasi-satellites had deep close-encounters very near to the escape velocity from the encounter distance.  The first three examples in Figure~\ref{FIG:5tau_nep} (top three rows) also help to demonstrate the diverse influence of the same secondary resonance on three different particles, as highlighted by the arrows in the left column.  This secondary resonance completely ejects particle A, moderately increases the libration amplitude of particle B, and sharply decreases the libration amplitude of particle C.  

\mar{Fig.~\ref{FIG:7tau_jup}}

Examples of escaping Jovian quasi-satellites are shown in Figure~\ref{FIG:7tau_jup}, from a simulation with $\tau = 10^6$ years.  In this simulation $\approx$80\% of the escaping Jovian quasi-satellites experienced a close-encounter with Jupiter within R$_\msct{H}$.  These encounters occurred just as the quasi-satellites were leaving the resonance.  As many as $\approx$20\% of Jovian quasi-satellites came within R$_\msct{H}$/4 as they left the resonance and $\approx$2.5\% come within R$_\msct{H}$/10.  Close-encounters of escaping Jovian quasi-satellites often occurred below the 2-body escape velocity from the encounter distance.  This is a consequence of solar gravity, which is still important inside R$_\msct{H}$, though not dominant.  These quasi-satellites are still traveling slightly faster than the
velocity needed to escape back outside R$_\msct{H}$, where solar
gravity can reclaim them.

Although not modeled here, we speculate that the flux of low-velocity quasi-satellites deep inside the planet's Hill radius could provide a source of objects capable of being captured as irregular satellites.   In our models there was no drag mechanism within the simulation to dissipate a test particle's energy (other than impacting the planet) so the quasi-satellites attained free heliocentric orbits after their close-encounters.  However, other workers (Philpott \etal 2010) have recently demonstrated that tidal stripping of a binary object passing deep within a planet's Hill radius could result in one member of the binary being retained as a captured irregular-type satellite while the escaping member carries away the excess energy.  Mann (2007) estimated that 6-10\% of Jupiter's current Trojan population is binary.  If quasi-satellites existed in similar proportions then during the late stages of planetary migration there could have been a steady influx of low-velocity binary quasi-satellites available for capture as irregular satellites.

\vspace{-0.5cm}
\section*{{\normalsize 5  \  \  DISCUSSION}}
\vspace{-0.2cm}

The experiments described in this paper provide several insights into quasi-satellite resonant trapping and orbital dynamics during planetary migration.  Some of our work suggests that there may be both an earlier phase of quasi-satellite trapping and that the effects of secondary resonances on quasi-satellites might be on-going.  There may also be implications for the known asymmetry in Jupiter's current population of Trojan asteroids.  Each of these points is discussed briefly below.

According to the Nice model, for a relatively brief period shortly after Jupiter and Saturn crossed their mutual 2:1 resonance the orbital eccentricities of all four giant planets would have been highly elevated (Tsiganis \etal 2005).  In previous modeling of quasi-satellite trapping (Kortenkamp, 2005) the planet's orbital eccentricity was closely linked with the resonant trapping efficiency.   All other parameters being equal, planets with larger orbital eccentricities were capable of trapping significantly more quasi-satellites.  At higher planetary eccentricities it also becomes possible to trap quasi-satellites with very low eccentricity.  This can be visualized in the rotating reference frame of Figure~\ref{FIG:ref_frames}{A}\@.  As the planet's eccentricity increases the tiny ellipse that marks its path in Figure~\ref{FIG:ref_frames}{A} gets larger.  The space inside this ellipse then becomes available for trapping of low eccentricity quasi-satellites.  During this stage it may have been possible for the planets to trap large populations of low eccentricity quasi-satellites.  The fate of this earlier population of quasi-satellites is uncertain.  As planetary eccentricity damped down the planet would have ``squeezed'' the low eccentricity quasi-satellites (in the point of view of Fig.~\ref{FIG:ref_frames}{A}).  This could lead to transition of quasi-satellites to higher eccentricity orbits, escape of the quasi-satellites, and close-encounters with the planet.  Furthermore, although the 2:1 resonance with Uranus is the dominant force acting on the evolution of Neptune Trojans for the initial conditions modeled here, weaker resonances may also play a role under differing initial conditions.  If Uranus and Neptune are started at a somewhat earlier stage (closer together) then secondary resonances associated with Uranus's 5:3, 7:4 and 9:5 became quite important in the evolution of Neptune Trojans (Kortenkamp \etal 2004).  In fact, with Uranus and Neptune starting a few AU closer together these resonances combine to eject a greater fraction of Neptune Trojans than the 2:1 alone.  This suggests that under somewhat different initial conditions --- with more tightly spaced planets on higher eccentricity orbits --- the efficiency of quasi-satellite trapping may have been significantly higher than in the present work.

In some of our simulations the exchange and escape of Trojans and quasi-satellites continues very late into the migration process, in some cases as late as 6-9$\tau$ (e.g., Figs.~\ref{FIG:1_4_planets_nep} \&~\ref{FIG:7tau_jup}).  At this stage migration is essentially complete ($>99.75$\% by $6\tau$) and the orbital evolution of the planets is dominated by their own mutual perturbations rather than the effects of the migration force.   Yet the secondary resonances continue to act.  This suggests that the process may be on-going and associated with the near-resonant configuration of some of the giant planets in our solar system, such as the Jupiter-Saturn near 2:5 resonance and the Uranus-Neptune near 2:1 resonance.  This could be one of the mechanisms responsible for destabilizing quasi-satellites of the giant planets on $10^8$ to $10^9$ year timescales in simulations performed by Wiegert \etaldot (2000).

Finally, a paper on the dynamical evolution of Trojans is not complete without some discussion of the leading--trailing asymmetry in the number of Jupiter's Trojan asteroids.  In the mid-1970s, following analysis of plates from the Palomar-Leiden Surveys (PLS), this asymmetry became quite dramatic, with 85 asteroids in Jupiter's leading \L{4} region versus just 15 in the trailing \L{5} region.   
Most workers recognized the possibility that the asymmetry was due entirely to observational selection effects, as the PLS disproportionally covered Jupiter's \L{4} region and the \L{5} region suffered from being both low in the Northern summer sky and having a dense galactic background (Shoemaker \etal 1989).  However, over the last 36 years Jupiter and its Trojan regions have completed three circuits across the sky and yet the asymmetry has been stubbornly persistent.  We have analyzed the number of Jupiter Trojans against their discovery dates using the full listing of Trojans maintained by the Minor Planet Center.  The population has only been in balance two times in the century since the first Trojan (588 Achilles) was discovered by Wolf (1906).  These were in 1907, when there were two Trojans known, and again in the mid-1930s when 10 were known.  Other than these times, a significant asymmetry in the \L{4}:\L{5} ratio of Trojans remained even as the number of Trojans grew by nearly three orders of magnitude, from 3 objects in 1908 (67:33); 15 in 1958 (67:33); 150 in 1985 (62:38); 1500 in 2002 (60:40); and 2300 in 2008 (54:46).  Recent analysis (Szab\'{o} \etal 2007) using the Sloan Digital Sky Survey Moving Object Catalog added 563 new Trojans and, controlling for observational selection effects, confirmed a real 60:40 asymmetry.

There has been much debate over the origin of this asymmetry and at one point the mystery even found its way into a popular work of science fiction (Clarke, 1993).  Drag forces acting on Trojan particles are known to act asymmetrically (Peale, 1993; Marzari \& Scholl, 1998; Kortenkamp \& Hamilton, 2001), but in the wrong sense, favoring trapping into the trailing \L{5} rather than the leading \L{4}.  Planetary migration is known to induce an asymmetrical trapping of objects into other resonances (Murray-Clay \& Chiang, 2005), so perhaps migration similarly influences Trojan trapping?   Modeling by Gomes (1998) hinted at such an effect.

\mar{Fig.~\ref{FIG:L4_to_L5}}

Most escaping Trojans in our simulations broke free from the 1:1 resonance region and obtained free heliocentric orbits.  However, Example F in Figure~\ref{FIG:qs_trapping_exp} highlights a potentially important effect.  In our simulations the secondary resonances were found to be capable of redistributing some Trojans from \L{4} into \L{5} and vice versa.  Figure~\ref{FIG:L4_to_L5} provides a more detailed demonstration of this effect.  In this example a leading \L{4} Neptune Trojan initially has a very stable low libration amplitude that appeared to be immune to the passage of the secondary resonances past the 4th and 3rd harmonics of \fnt{1:1}.  This Trojan was then relatively quickly transformed into a trailing \L{5} Trojan when the 2nd harmonic was reached (\fun{2:1} = 2 \fnt{1:1}).   Witnessing the redistribution of Trojans in our simulations raises the question as to whether this effect is symmetrical with respect to \L{4} and \L{5}.  

We searched the output from simulations involving 14,000 initial Jupiter Trojan test particles equally distributed between the leading \L{4} and the trailing \L{5} and with a migration timescale of $\tau = 10^7$ years.  These simulations started with more compact planetary orbits than shown in Table~~\ref{tab:model_parameters}, but with Jupiter and Saturn already past their mutual 2:1 resonance.  Over the course of $5\tau$ we observed 53 Jupiter Trojans switching from \L{4} to \L{5} and 55 undergoing the reverse transition, from \L{5} to \L{4}.  Furthermore, after $5\tau$ there were 140 Trojans remaining in \L{4} and only one of these had initially been in \L{5}.  On the other side, there were 146 \L{5} survivors, 2 of which were initially in \L{4}.  None of the survivors had switched more than once.  For Neptune Trojans with the same migration timescale of $\tau = 10^7$ years we found many hundreds of Trojans switching repeatedly back and forth between \L{4} and \L{5}.  Only a few tens of Trojans survived to $5\tau$ and there were 7 in \L{4} that were initially in \L{5}, while 5 were in \L{5} that were initially in \L{4}.  Based on these simulations it appears the effect is symmetrical, although we did not study a range in initial planetary orbits nor a wide range in migration rates.  More work is needed to determine if a faster planetary migration rate can influence the \L{4}:\L{5} symmetry in a manner similar to but in the opposite sense as what Peale (1993) showed takes place for strong drag forces acting on planetesimals.

\vspace{-0.5cm}
\section*{{\normalsize 6  \  \  ACKNOWLEDGMENTS}}
\vspace{-0.2cm}

This material is based upon work supported by the National Science Foundation under grant AST0607777 and the National Aeronautics and Space Administration under grant NNX10AJ61G.

\vspace{-0.5cm}
\section*{{\normalsize 7  \  \  APPENDIX --- HISTORICAL NOTE ON NOMENCLATURE}}
\vspace{-0.2cm}

There has been some ambiguity in the use of the
descriptors \L{4} and \L{5} for the Lagrangian equilibrium points.  
Charlier (1906) used \L{4} to indicate
the leading Lagrange point about which 1906\,TG (588 Achilles) is librating, citing
his extensive published lectures (Charlier, 1902) to support his case.
A short time later, when two more asteroids were added to the
``Jupiter Group,'' Charlier (1907) reversed his convention and began
using \L{4} to indicate the trailing point.  In parenthetical comments
he mentioned that his earlier paper was in error.  Unfortunately,
Charlier's efforts to correct himself came too late, the damage had
already been done.  In fact, on the very same page of
\textsl{Astronomische Nachrichten} where Charlier (1907) attempts to
reverse his notation, referring to \L{4} and \L{5} as the trailing and
leading points, Str\"omgren (1907) uses the opposite convention,
presumably following Charlier's 1906\,TG paper.  A footnote added to
Str\"omgren's paper by the editor highlights the different notation
used by the two authors.  In this paper we follow Charlier's (1906)
original designation and recent custom (e.g., Murray and
Dermott, 1999), referring to \L{4} and \L{5} as the leading and
trailing points, respectively.  While there is considerable historical
momentum behind this convention it is not universally followed.
Therefore, we also try to write ``leading \L{4}'' and ``trailing
\L{5}'' wherever it is reasonable to do so.  The descriptors \L{1},
\L{2}, and \L{3} refer to the three unstable equilibrium points that
fall on the line formed by the Sun and the host planet.


\newcommand{\vpy}[4]{\textbf{#1}, #2-#3.}

\renewcommand{\aj}[4]{\textsl{Astron.\ J.} \vpy{#1}{#2}{#3}{#4}}
\newcommand{\an}[4]{\textsl{Astronomische Nachrichten} \vpy{#1}{#2}{#3}{#4}}
\newcommand{\ana}[4]{\textsl{Astron.\ Astrophys.\ }\vpy{#1}{#2}{#3}{#4}}
\renewcommand{\apj}[4]{\textsl{Astrophys.\ J.} \vpy{#1}{#2}{#3}{#4}}
\newcommand{\apjip}{\textsl{Astrophys.\ J.} (in press).}
\newcommand{\apjsub}{\textsl{Astrophys.\ J.} (submitted).}
\renewcommand{\apjl}[4]{\textsl{Astrophys.\ J. Lett.\ }\vpy{#1}{#2}{#3}{#4}}
\renewcommand{\apss}[4]{\textsl{Astrophys.\ Space Sci.\ }\vpy{#1}{#2}{#3}{#4}}
\renewcommand{\araa}[4]{\textsl{Annu.\ Rev.\ Astron.\ Astrophys.\ }\vpy{#1}{#2}{#3}{#4}}
\newcommand{\asl}[4]{\textsl{Astronomy Lett.\ }\vpy{#1}{#2}{#3}{#4}}
\newcommand{\cmda}[4]{\textsl{Celes.\ Mech.\ Dynamic.\ Astron.\ }\vpy{#1}{#2}{#3}{#4}}
\renewcommand{\icarus}[4]{\textsl{Icarus} \vpy{#1}{#2}{#3}{#4}}
\newcommand{\icarusip}[4]{\textsl{Icarus} (in press).}
\newcommand{\maps}[4]{\textsl{Meteor.\ Planet.\ Sci.\ }\vpy{#1}{#2}{#3}{#4}}
\renewcommand{\mnras}[4]{\textsl{Mon.\ Not.\ Royal Astro.\ Soc.\ }\vpy{#1}{#2}{#3}{#4}}
\newcommand{\moonpl}[4]{\textsl{Moon Planets} \vpy{#1}{#2}{#3}{#4}}
\newcommand{\nature}[4]{\textsl{Nature} \vpy{#1}{#2}{#3}{#4}}
\newcommand{\lpsc}[3]{#1 \textsl{Lunar Planet.\ Sci.\ Conf.}, abstract \##2 (#3).}
\newcommand{\lpscpp}[4]{\textsl{Lunar Planet.\ Sci.\ Conf.\ }{\bf #1}, #2-#3 (abstract).}
\renewcommand{\pasp}[4]{\textsl{Pub.\ Astron.\ Soc.\ Pacific} \vpy{#1}{#2}{#3}{#4}}
\newcommand{\paz}[4]{\textsl{Pis'ma v.~Astronomicheskii Zhurnal} \vpy{#1}{#2}{#3}{#4}}
\renewcommand{\phd}[3]{\textsl{Ph.D. dissertation}, #1.}
\newcommand{\pia}[4]{\textsl{Proc.\ Irish Acad.\ }\vpy{#1}{#2}{#3}{#4}}
\newcommand{\pnas}[4]{\textsl{Proc.\ Natl.\ Acad.\ Sci.\ }\vpy{#1}{#2}{#3}{#4}}
\newcommand{\pss}[4]{\textsl{Plan.\ Spa.\ Sci.\ }\vpy{#1}{#2}{#3}{#4}}
\newcommand{\ptp}[4]{\textsl{Prog.\ Theoret.\ Phys.\ }\vpy{#1}{#2}{#3}{#4}}
\newcommand{\ptps}[4]{\textsl{Prog.\ Theoret.\ Phys.\ Supp.\ }\vpy{#1}{#2}{#3}{#4}}
\newcommand{\science}[4]{\textsl{Science} \vpy{#1}{#2}{#3}{#4}}
\newcommand{\ssrsub}{\textsl{Space Sci.\ Reviews} (submitted).}

\newcommand{\astii}[2]{In \textsl{Asteroids II}, 
			(R.P. Binzel, T. Gehrels, and M.S. Matthews, Eds.),
			pp.~#1--#2. Univ.\ Arizona Press, Tucson.}

\newcommand{\mars}[2]{In \textsl{Mars}
        (H.H. Kieffer, B.M. Jakosky, C.W. Snyder, and M.S. Matthews, Eds.)
        pp.~#1-#2. Univ.\ Arizona Press, Tucson.}

\newcommand{\oem}[2]{In \textsl{Origin of the Earth and Moon}
        (R.M. Canup, K. Righter, Eds.)
        pp.~#1-#2. Univ.\ Arizona Press, Tucson.}

\newcommand{\ppiii}[2]{In \textsl{Protostars and Planets III}
        (E.H. Levy and J.I. Lunine, Eds.) 
        pp.~#1-#2. Univ.\ Arizona Press, Tucson.}

\newcommand{\ppiv}[2]{In \textsl{Protostars and Planets IV}
        (V. Mannings, A.P. Boss, and S.S. Russell, Eds.)
        pp.~#1-#2. Univ.\ Arizona Press, Tucson.}

\newcommand{\soj}[2]{In \textsl{Satellites of Jupiter}
        (D. Morrison, Ed.)
        pp.~#1-#2. Univ.\ Arizona Press, Tucson.}

\newcommand{\nep}[2]{In \textsl{Neptune and Triton},
        (D.P. Cruikshank, Ed.)
        pp.~#1-#2. Univ.\ Arizona Press, Tucson.}


\vspace{-0.5cm}
\section*{{ \normalsize 8  \  \  REFERENCES}}
\vspace{-0.2cm}

\begin{list}{}{\setlength{\leftmargin}{0.5cm} 
                \setlength{\rightmargin}{0cm} 
                \setlength{\itemindent}{-0.5cm}}

\item {\sc Brown, E.W.},  1911.
        On a new family of periodic orbits in the
        problem of three bodies.  
        \mnras{71}{438}{454}{1911}

\item {\sc Chambers, J.E.}, 1999.
	A hybrid symplectic integrator that permits 
	close encounters between massive bodies.  
        \mnras{304}{793}{799}{1999}

\item 	{\sc Charlier, C.V.L.}, 1902.
	\textsl{Die Mechanik des Himmels.}
	Verlag von Veit, Leipzig.

\item 	{\sc Charlier, C.V.L.}, 1906.
	\"Uber den planeten 1906\,TG.
	\an{171}{213}{216}{1906}

\item 	{\sc Charlier, C.V.L.}, 1907.
	\"Uber die bahnen der planeten (588) [1906\,TG], 1906\,VY 
 	und 1907\,XM.
	\an{175}{89}{90}{1907}

\item {\sc Clarke, A.C.}, 1993.
      The Hammer of God.
      Bantum Books, New York.
      
\item {\sc Connors, M., Chodas, P., Mikkola, S., Wiegert, P., Veillet, C., and Innanen, K.}, 2002.
        Discovery of an asteroid and quasi-satellite in an Earth-like
        horseshoe orbit.
        \maps{37}{1435}{1441}{2002}

\item {\sc Connors, M., Veillet, C., Brasser, R., Wiegert, P.,
        Chodas, P., Mikkola, S., and Innanen, K.}, 2004.
        Discovery of Earth's quasi-satellite.
        \maps{39}{1251}{1255}{2004}

\item {\sc Gomes, R.S.}, 1998.
        Dynamical effects of planetary migration on
        primordial Trojan-type asteroids. 
        \aj{116}{2590}{2597}{1998}


\item {\sc Gomes, R.S., Tsiganis, K., Morbidelli, A., and Levison, H.F.}, 2005.
      Origin of the cataclysmic Late Heavy Bombardment period 
      of the terrestrial planets.
        \nature{435}{466}{469}{2005}

\item {\sc Hahn, J.M. and Malhotra, R.},  1999.
        Orbital evolution of planets embedded in a planetesimal disk. 
        \aj{117}{3041}{3053}{1999}

\item {\sc H\'enon, M.}, 1969.  
        Numerical exploration of the restricted problem, V.  
        \ana{1}{223}{238}{1969}

\item {\sc Jackson, J.},  1913.
        Retrograde satellite orbits.
        \mnras{74}{62}{82}{1913}

  \item {\sc Kortenkamp, S.J.,} 2005.
        {An efficient, low-velocity, resonant mechanism for
        capture of satellites by a protoplanet.}
        \icarus{175}{409}{418}{2005}



  \item {\sc Kortenkamp, S.J. and Wetherill, G.W.,} 2000.
        {Terrestrial planet and asteroid formation in the presence of 
        giant planets I.\ \ Relative velocities of planetesimals subject 
        to Jupiter and Saturn perturbations.} 
        \icarus{143}{60}{73}{2000}

\item   {\sc Kortenkamp, S.J. and Hamilton, D.P.}, 2001.
        {Capture of Trojan asteroids in the early solar nebula}, abstract 25.06, 
        $\mathrm{33^{rd}}$ DPS Meeting, New Orleans, LA.

  \item {\sc Kortenkamp, S.J., Wetherill, G.W., and Inaba, S.,} 2001.
        {Runaway growth of planetary embryos facilitated by 
        massive bodies in a protoplanetary disk.}
        \science{293}{1127}{1129}{2001}

  \item {\sc Kortenkamp, S.J., Malhotra, R., and Michtchenko, T.,} 2004.
        {Survival of Trojan-type companions of Neptune 
        during primordial planet migration.}
        \icarus{167}{347}{359}{2004}

\item {\sc Levison, H.F. and Duncan, M.J.}, 1994.
	The long-term dynamical behavior of short-period comets.
        \icarus{108}{18}{36}{1994}

\item {\sc Levison, H.F., Morbidelli, A., Vanlaerhoven, C., Gomes, R., and Tsiganis, K.}, 2008.
	Origin of the structure of the Kuiper belt during a dynamical instability in the orbits of Uranus and Neptune.
        \icarus{196}{258}{273}{2008}

\item {\sc Michtchenko, T.A., Beaug\'e, C., and Roig, F.,} 2001.
	Planetary migration and the effects of mean motion resonances  on Jupiter's Trojan asteroids.  
	\aj{122}{3485}{3491}{2001}

\item {\sc Mann, R.K., Jewitt, D., and Lacerda, P.,} 2007.
	Fraction of contact binary Trojan asteroids.  
	\aj{134}{1133}{1144}{2007}
 
\item {\sc Malhotra, R.,} 1995.
	The origin of Pluto's orbit: Implications for the solar system beyond Neptune.
	\aj{110}{420}{429}{1995}

\item 	{\sc Marzari, F., and Scholl, H.} 1998.  
	Capture of Trojans by a growing proto-Jupiter. 
	\icarus{131}{41}{51}{1998}

\item {\sc Morbidelli, A., Levison, H.F., Tsiganis, K., and Gomes, R.}, 2005.
      Chaotic capture of Jupiter's Trojan asteroids in the early Solar System.
      \nature{435}{462}{465}{2005}

\item {\sc Murray-Clay, R.A. and Chiang, E.I.}, 2005.
	A signature of planetary migration: The origin of asymmetric capture in the 2:1 resonance.
	\apj{619}{623}{638}{2005} 

\item {\sc Murray, C. and Dermott, S.F.}, 1999.
      \textsl{Solar System Dynamics}, Cambridge Univ.\ Press, Cambridge.

\item {\sc Namouni, F.}, 1999.
        Secular interactions of coorbiting objects.
        \icarus{137}{293}{314}{1999}
        
\item {\sc Nesvorn\'y, D., Thomas, F., Ferraz-Mello, S., and Morbidelli, A.}, 2002.
        A perturbative treatment of the co-orbital motion.
        \cmda{82}{323}{361}{2002}
        
\item {\sc Nesvorn\'y, D. and Vokrouhlick\'y, D.}, 2009.
        Chaotic capture of Neptune Trojans.
        \aj{137}{5003}{5011}{2009}

\item 	{\sc Peale, S.J.}, 1993.
	The effect of the nebula on the Trojan precursors.  
	\icarus{106}{308}{322}{1993}

\item {\sc Philpott, C.M., Hamilton, D.P., and Agnor, C.B.}, 2010.
	  Three-body capture of irregular satellites: Application to Jupiter.
	  \icarus{208}{824}{836}{2010}

\item {\sc Peterson, C.A.}, 1976.
        Analysis of some solar system dynamics problems.
        \phd{Massachusetts Institute of Technology}{1976}{}

\item 	{\sc Shoemaker, E.M., Shoemaker, C.S., and Wolfe, R.F.}, 1989. 
	Trojan asteroids: Population, dynamical 
	structure, and origin of the L4 and L5 swarms.  
	\astii{487}{523}

\item {\sc Stoer, J. and Bulirsch, R.}, 1980.
	\textsl{Introduction to Numerical Analysis},
	Springer-Verlag, NY.

\item 	{\sc Str\"omgren, E.}, 1907.
	Zer entdeckung eines dritten kleinen planeten der 
	Jupitergruppe 1906\,VY.
	\an{175}{89}{90}{1907}

\item 	{\sc Szab\'o, Gy.M., Ivezi\'c, \v{Z}., Juri\'c, M., and Lupton, R.}, 2007.
	The properties of Jovian Trojan asteroids listed in SDSS Moving Object Catalogue 3.
	\mnras{377}{1393}{1406}{2007}

\item {\sc Tsiganis, K., Gomes, R., Morbidelli, A., and Levison, H.F.}, 2005.
      Origin of the orbital architecture of the giant 
      planets of the Solar System.
      \nature{435}{459}{461}{2005}
	
\item {\sc Wajer, P.}, 2009.
	2002 AA29: Earth's recurrent quasi-satellite?
	\icarus{200}{147}{153}{2009}
	
\item {\sc Wajer, P.}, 2010.
	Dynamical evolution of EarthÕs quasi-satellites: 2004 GU9 and 2006 FV35.
	\icarus{209}{488}{493}{2010}
        
\item {\sc Wiegert, P., Innanen, K., Mikkola, S.}, 2000.
        The stability of quasi-satellites in the outer solar system.
        \aj{119}{1978}{1984}{2000}
	
\item {\sc Wiegert, P., Connors, M., Brasser, R., Mikkola, S., Stacey, G., and Innanen, K.}, 2005.
	Sleeping with an elephant: Asteroids that share a planet's orbit (abstract).
	{\em Journal of the Royal Astronom.\ Soc.\ Canada} {\bf 99}, 145.

\item {\sc Wisdom, J. and Holman, M.}, 1991.
	Symplectic maps for the N-body problem.
        \aj{102}{1528}{1538}{1991}

\item 	{\sc Wolf, M.}, 1906.
	Photographische aufnahmen von kleinen planeten.
	\an{170}{353}{354}{1906}

\end{list}

\mylabel{last_ms_page}

\newpage
  
\newcounter{tabs}
   
\vspace*{3cm}
   
\renewcommand{\footnoterule}{\vspace{-0.2cm}}
\refstepcounter{tabs}
\label{tab:model_parameters}
\label{last_table}
\begin{minipage}{15cm}
\small
 \begin{center}
 \begin{tabular}{c@{\hspace{1cm}}ccc}
 \multicolumn{4}{c} {\bf Table~\ref{tab:model_parameters}} \\
 \multicolumn{4}{c} {Nominal Initial Heliocentric Planetary Configuration\footnote[2]{In units of solar masses, AUs, and degrees referred to invariable plane and J2000 mean equinox.}} \\[0.2cm] 
 \hline \hline \\[-0.2cm] 
  Planet	& Mass \\
		& Semi-Major Axis & Eccentricity & Inclination  \\
		& Long.~Ascend.~Node & Arg.~Pericenter & Mean Anomoly  \\[0.2cm] 
 \hline \\[-0.2cm] 
  Jupiter &  9.5479E$-$04 \\ 
  	  &  5.2759E$+$00   &   3.5909E$-$02   &   2.8040E$-$01  \\
   	  &  7.7659E$+$01   &   7.7871E$+$01  &   1.9062E$+$01\\[0.5cm]
  Saturn  & 2.8559E$-$04 \\ 
  	  &  9.2631E$+$00   &   6.7107E$-$02   &   9.9580E$-$01  \\
   	  &  2.9932E$+$02   &   2.8308E$+$02  &   1.6736E$+$02\\[0.5cm]
  Uranus  & 4.3728E$-$05 \\ 
  	  &  1.7965E$+$01   &   5.8528E$-$02   &   9.1650E$-$01  \\
   	  &  2.0099E$+$02   &   1.2073E$+$02  &   9.6331E$+$01\\[0.5cm]
  Neptune & 5.1776E$-$05 \\ 
  	  &  2.7041E$+$01   &   1.0125E$-$02  &   7.3140E$-$01  \\
   	  &  2.1892E$+$02   &   2.3783E$+$02  &  9.4434E$+$01\\[0.5cm]
  \hline \hline
 \end{tabular}
 \end{center}
\normalsize
\end{minipage}

\clearpage






	\gdef\CAPTIONrefframes{
	  The orbits of a planet and  a
	  resonant quasi-satellite are shown projected onto
	  the {\sf X-Y} plane in two different reference frames;
	  ({\sf A}) a sun-centered frame rotating with the mean
	  orbital motion of the planet, ({\sf B}) a
	  planet-centered frame with the inertial orientation of the {\sf X-Y} axes preserved (i.e., the sun circulates around the planet in {\sf B}, but stars are fixed).  In the rotating frame of {\sf A}, the orbital eccentricities of the planet and the quasi-satellite cause
	  both objects to follow elliptical paths.  Three of the traditional Lagrange equilibrium
	  points, {\sf \L{3}, \L{4}} and {\sf \L{5}}, are indicated.
	  In the geocentric frame of {\sf B} the
	  quasi-satellite traces out the double-lobed looping path once during each orbit around the sun (the quasi-satellite remains in the same quadrant of the sky as seen from the planet).  Over much longer time frames the looping path followed by the quasi-satellite slowly precesses clockwise around the planet, corresponding to retrograde motion as seen from the planet. In {\sf B} the size of the planet's Hill sphere (Neptune's in this example) is indicated by the bold dot at the origin.  Note that in this example the quasi-satellite remains more than 10 Hill radii away from the planet at all times.}



	\gdef\CAPTIONplanetaryorbits{
	   Four panels showing examples of the evolution
	with time of the semi-major axes of the four giant planets in
	a representative migration simulation beginning with the nominal initial conditions given in Table~\ref{tab:model_parameters}.  
	  The planets were subject to mutual
	gravitational perturbations and a drag force which caused
	their orbits to migrate---Jupiter inward; Saturn, Uranus, and
	Neptune outward. 	   Time is expressed
	in units of $\tau$, the characteristic migration timescale.
	After a time of $5\tau$ migration is 99.33\% complete and
	subsequent orbital evolution is dominated by mutual planetary
	gravitational perturbations rather than the migration force.
	The grey lines indicate the locations of the outer 2:1 mean-motion resonance with Uranus (top panel) and the inner 2:5 mean-motion resonance with Saturn (bottom panel).}



	\gdef\CAPTIONinitialtrojans{
	  The initial population of Trojan companions is shown projected
        onto the {\sf X-Y} plane of the same sun-centered rotating reference frame
        as used in Figure~\ref{FIG:ref_frames}{\sf A}\@.  Black
        points mark the initial positions of 5000 leading  \L{4} and 5000 trailing \L{5} Trojans generated from the orbital element distribution of Jupiter's known Trojan asteroids. }



	\gdef\CAPTIONqshires{
	  A high-resolution plot showing libration of the critical argument, $\phi_\msct{1:1}$, of an object trapped in Neptune's quasi-satellite resonance (top panel).  The quasi-satellite's libration with respect to Neptune displays two dominant frequencies, as indicated by the power spectrum of $\phi_\msct{1:1}$ (bottom panel).  A narrow peak at about 6 kyr$^{-1}$ corresponds with the orbital period around the sun (about 166 years for the starting conditions of this particular migration simulation).  At a much slower frequency a broad peak in the spectrum of $\phi_\msct{1:1}$ is roughly commensurate with the circulation frequency, $f^\msct{UN}_\msct{2:1}$, of the Uranus-Neptune near 2:1 resonance.  When $\phi_\msct{1:1}$ passes through $0^\circ$ (see top panel near times of 2.0755, 2.077, 2.078, etc.) the two lobes in Figure~\ref{FIG:ref_frames}{B} are of equal size. }
 


	\gdef\CAPTIONnewejection{
	  The top panel shows evolution of the critical argument, $\phi_\msct{1:1}=\lambda-\lambda_\msct{Nep}$, for a leading \L{4} Trojan in a simulation with migration timescale $\tau = 10^6$ years.  For the bottom two panels a fast Fourier transform (FFT) was used to obtain power spectra of $\phi_\msct{1:1}$ (15 spectra in middle panel) and spectra of the critical argument of the Uranus-Neptune 2:1 near-resonance ($\phi^\msct{UN}_\msct{2:1}=2\lambda_\msct{Nep} - \lambda_\msct{Ura} - \tilde{\omega}_\msct{Ura}$, 15 spectra in bottom panel).  Each FFT used 2048 points sampled every 50 years (see FFT interval bars).
	  Spectra of $\phi^\msct{UN}_\msct{2:1}$ are shown in linear units of power while spectra of $\phi_\msct{1:1}$ are shown in units of log power in order to simultaneously resolve the fundamental frequency (\fnt{1:1}) and its higher harmonics.  Each spectrum is individually normalized. As \fun{2:1}  slows (spectra in bottom panel) it first passes the $4^\msc{th}$ and then the $3^\msc{rd}$ harmonic of \fnt{1:1} (spectra in middle panel).  Coincident with each of these passages the Trojan experiences a change in the libration amplitude of $\phi_\msct{1:1}$ (top panel).  When the $2^\msc{nd}$ harmonic of \fun{2:1} reaches the $5^\msc{th}$ harmonic of \fun{2:1} the Trojan is temporarily trapped in this 2:5 secondary resonance.  This secondary resonance drives the Trojan out of the leading \L{4} region and into the 1:1 quasi-satellite resonance with Neptune.}



	\gdef\CAPTIONqstrappingexp{
	  Six examples demonstrating the diversity of quasi-satellite resonant trapping.  Some quasi-satellites were initially leading \L{4} or trailing \L{5} Trojans (A \& B, respectively) before secondary resonances forced them into the quasi-satellite region.  Occasionally, L4 and L5 Trojans first had their libration amplitudes enlarged into the horseshoe region (C \& D, respectively), before subsequent waves of secondary resonances shrank their libration amplitudes at precisely the time needed to inject them into the quasi-satellite resonance.  Example C additionally demonstrates the circuitous path some objects can follow before reaching the quasi-satellite resonance --- \L{4} to horseshoe, briefly back to \L{4}, back to horseshoe, brief escape to free orbit, back to horseshoe, and finally into the quasi-satellite resonance.  Secondary resonances are also capable of trapping free objects directly into a quasi-satellite state (E). These effects sometimes combined in a number of ways, as shown in Example F.  Here an L5 Trojan is transferred first into the quasi-satellite resonance, then into L4, and finally returned to its initial, L5, state.}


%

	\gdef\CAPTIONonefourplanetexp{
	  These panels show the evolution of representative Neptune quasi-satellites in three different migration simulations, all with the same migration timescale ($\tau = 10^6$ years) and all with identical initial populations of quasi-satellites.  The three simulations involved Neptune migrating alone (top row), Jupiter, Saturn, and Neptune, but no Uranus (middle row), and the nominal model with all four giant planets (bottom row).  The columns show the resonant argument $\phi_\msct{1:1}$ (left column), distance from Neptune during each close-encounter (middle column), and the velocity with respect to Neptune at the close-encounter distance (right column).  By definition, a close-encounter occurs once during each orbit around the sun as the quasi-satellite follows the looping path shown in 
Fig.~\ref{FIG:ref_frames}.  Secondary resonances associated with Uranus and Neptune approaching their mutual 2:1 resonance (as in the example shown in Figure~ \ref{FIG:new_ejection}) cause variations in the libration amplitude of $\phi_\msct{1:1}$ (bottom row, left column).  Larger libration amplitudes bring quasi-satellites relatively closer to Neptune (bottom row, middle column) at relatively lower encounter velocities (bottom row, right column).}



	\gdef\CAPTIONfivetaunep{
		Five examples demonstrating the escape of Neptune quasi-satellites during the stage of planetary migration between  2 and 5$\tau$.  The simulation included all four giant planets and had a value of $\tau = 10^7$ years.   The columns show the resonant argument ($\phi_\msct{1:1}$, left column), distance from Neptune in Hill radii, R$_\msct{H}$, during each close-encounter (D$_\msct{ce}$, middle column), and the velocity with respect to Neptune at the close-encounter distance (right column).  Filled grey circles indicate distances and corresponding velocities for quasi-satellites that come within R$_\msct{H}$.  Single open grey circles further indicate data points for encounters within a quarter R$_\msct{H}$ while double open grey circles highlight encounters at less than a tenth of R$_\msct{H}$.  The 2-body escape velocity from each of these distances from Neptune is indicated in the right column.  
		In Examples A-C arrows in the first column highlight the diverse effects of the same secondary resonance on three different quasi-satellites, triggering an escape (A), a slight de-stabilization of the libration amplitude (B), or a sharp stabilization of the libration amplitude (C).}

%

	\gdef\CAPTIONseventaujup{
		Similar to Figure~\ref{FIG:5tau_nep} except for objects escaping Jupiter's quasi-satellite region between 0 and 7$\tau$.  The simulation included all four giant planets and had a value of $\tau = 10^6$ years.   
		In Examples A and B the escaping quasi-satellites have close encounters within R$_\msct{H}/10$ of Jupiter while traveling significantly below the 2-body escape velocity from this distance.
			}

%
%

	\gdef\CAPTIONfourtofive{
        Evolution of an initial leading \L{4} Trojan shown with the same formatting as used in Figure~\ref{FIG:new_ejection}. 
As \fun{2:1}  slows (spectra in bottom panel) it approaches the $2^\msc{nd}$ harmonic of \fnt{1:1} (spectra in middle panel).  When  \fun{2:1} hits the second harmonic of \fnt{1:1} the Trojan begins to experience dramatic changes in the libration amplitude of $\phi_\msct{1:1}$ (top panel).  The two frequencies remain locked as this secondary resonance drives the object out of the leading \L{4} region and into the trailing \L{5} region. This redistribution from  \L{4} to \L{5} also allows the Trojan to decouple from the secondary resonance, becoming a relatively stable \L{5} Trojan.  Small arrows in the middle panel indicate the effect that \fun{2:1} has on \fnt{1:1} as Uranus and Neptune converge upon their mutual 2:1 resonance.}

	




\newcounter{figs}

\newpage

\noindent
\begin{minipage}[t][15.0cm][t]{17cm}   \refstepcounter{figs}  \label{FIG:ref_frames}
        \sloppy 
        \vspace{-7cm}
        \centerline{  
          \hspace{-0cm}
          \scalebox{1.0}{ \includegraphics*[bb=0 0 550 550]{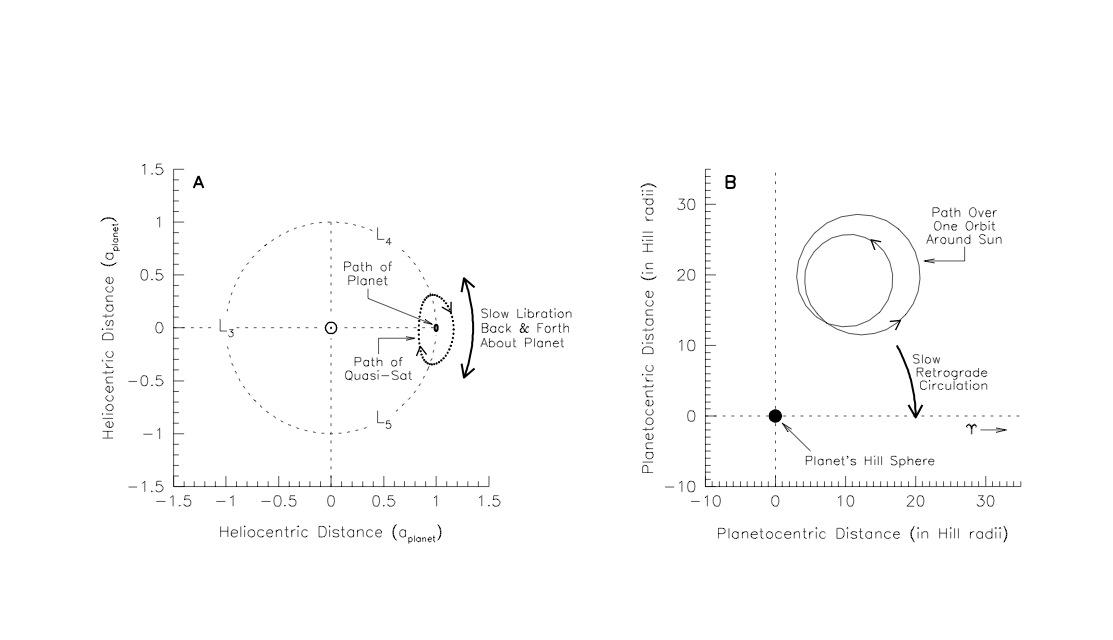}}}  
        \parbox[10cm]{16.5cm}{
        \renewcommand{\baselinestretch}{1.0}
        \mysize \sf \makebox[0cm]{}\\[-0cm]
        Figure~\ref{FIG:ref_frames}: 
	\CAPTIONrefframes
        }     
\end{minipage} 
\clearpage

\noindent
\begin{minipage}[t][15.0cm][t]{17cm}    \refstepcounter{figs} \label{FIG:planetary_orbits}
        \sloppy 
        \vspace{-5cm}
        \centerline{  
          \hspace{-1cm}
          \scalebox{1.0}{ \includegraphics*[bb=0 0 550 550]{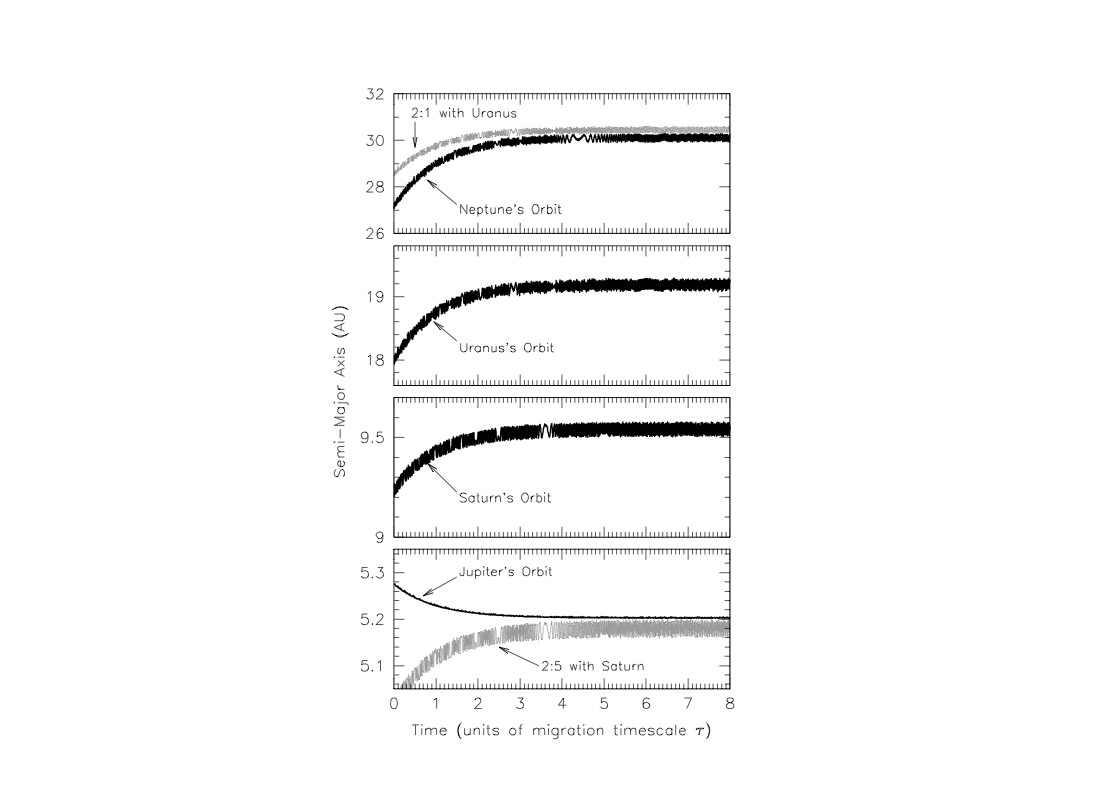}}}  
        \parbox[10cm]{16.5cm}{
        \renewcommand{\baselinestretch}{1.0}
        \mysize \sf \makebox[0cm]{}\\[-0cm]
        Figure~\ref{FIG:planetary_orbits}: 
	\CAPTIONplanetaryorbits
        }     
\end{minipage} 
\clearpage

\noindent
\begin{minipage}[t][17.0cm][t]{17cm}    \refstepcounter{figs} \label{FIG:initial_trojans}
        \sloppy 
        \vspace{-5cm}
        \centerline{  
          \hspace{-1cm}
          \scalebox{1.0}{ \includegraphics*[bb=0 0 550 550]{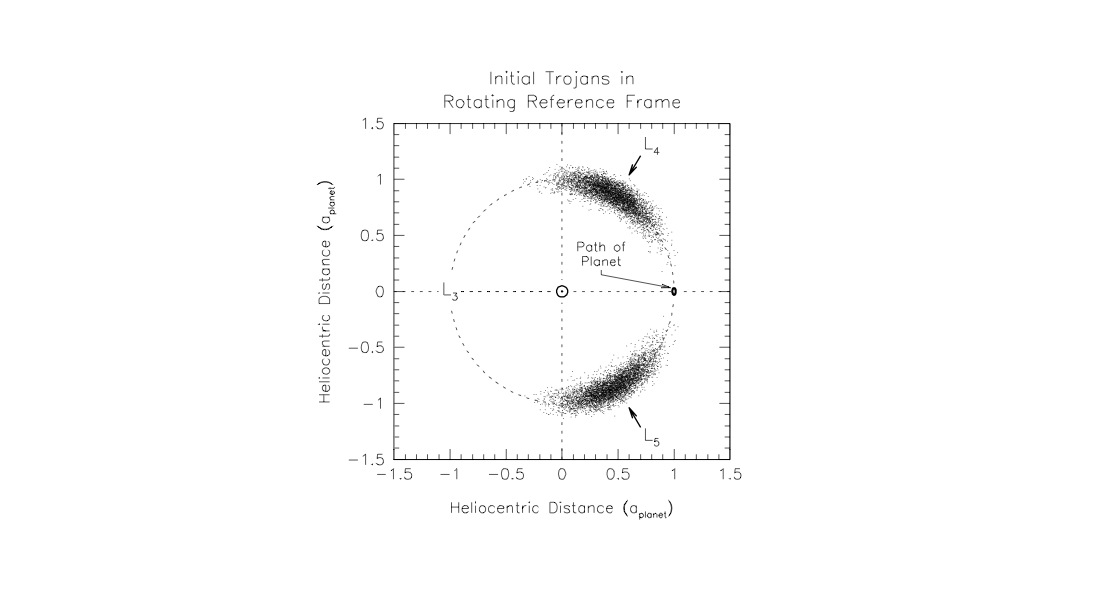}}}  
        \parbox[12cm]{16.5cm}{
        \renewcommand{\baselinestretch}{1.0}
        \mysize \sf \makebox[0cm]{}\\[-0cm]
        Figure~\ref{FIG:initial_trojans}: 
	\CAPTIONinitialtrojans
        }     
\end{minipage} 
\clearpage

\noindent
\begin{minipage}[t][15.0cm][t]{17cm}    \refstepcounter{figs} \label{FIG:qs_hires}
        \sloppy 
        \vspace{-2cm}
        \centerline{  
          \hspace{-1cm}
          \scalebox{0.8}{ \includegraphics*[bb=0 0 550 550]{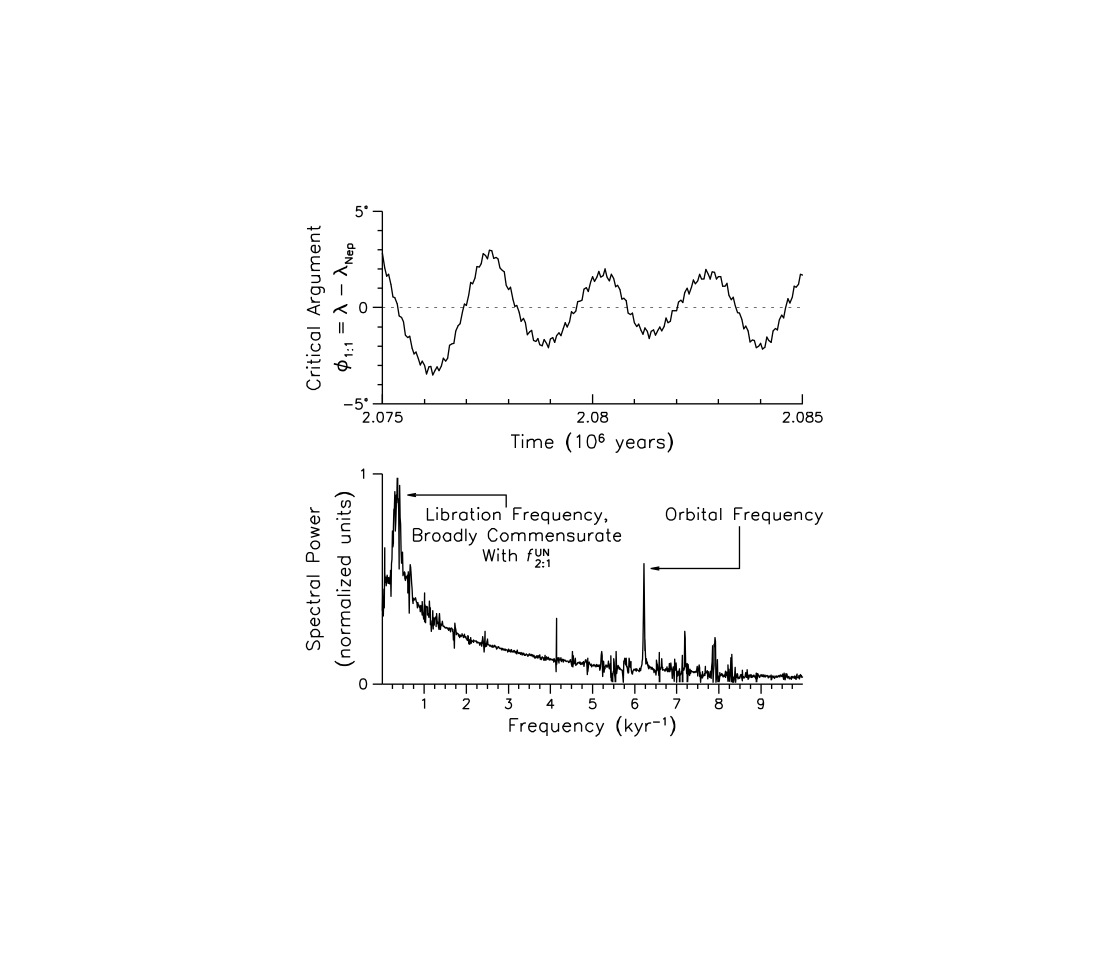}}}  
        \parbox[10cm]{16.5cm}{
        \renewcommand{\baselinestretch}{1.0}
        \mysize \sf \makebox[0cm]{}\\[-2cm]
        Figure~\ref{FIG:qs_hires}: 
	\CAPTIONqshires
        }     
\end{minipage} 
\clearpage

\noindent
\begin{minipage}[t][15.0cm][t]{17cm}    \refstepcounter{figs} \label{FIG:new_ejection}
        \sloppy 
        \vspace{-0cm}
        \centerline{  
          \hspace{-1cm}
          \scalebox{0.8}{ \includegraphics*[bb=0 0 550 575]{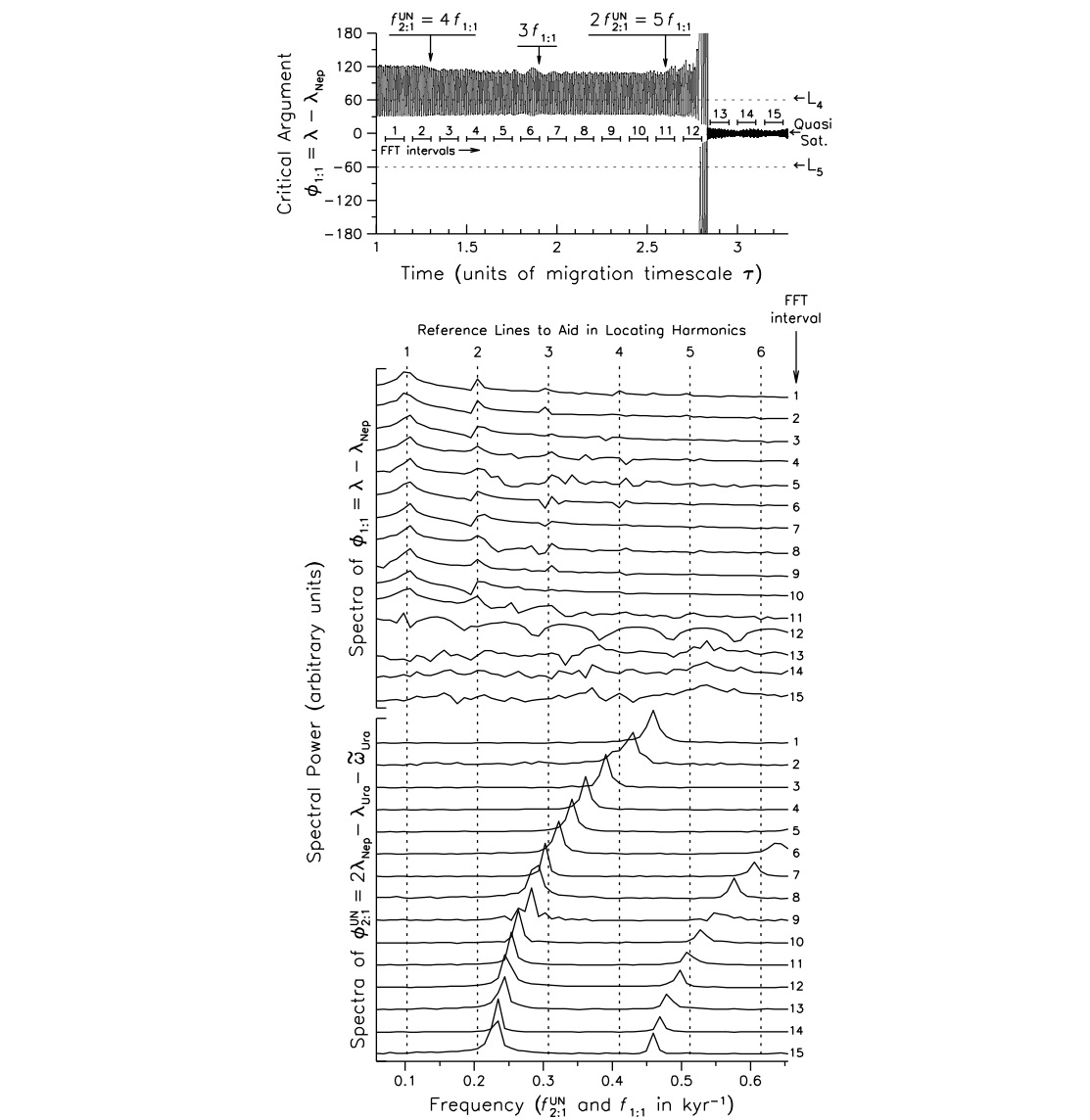}}}  
        \parbox[10cm]{16.5cm}{
        \renewcommand{\baselinestretch}{1.0}
        \mysize \sf \makebox[0cm]{}\\[0.25cm]
        Figure~\ref{FIG:new_ejection}: 
	\CAPTIONnewejection
        }     
\end{minipage} 
\clearpage

\noindent
\begin{minipage}[t][15.0cm][t]{17cm}    \refstepcounter{figs} \label{FIG:qs_trapping_exp}
        \sloppy 
        \vspace{-2cm}
        \centerline{  
          \hspace{-1cm}
          \scalebox{0.8}{ \includegraphics*[bb=0 0 550 550]{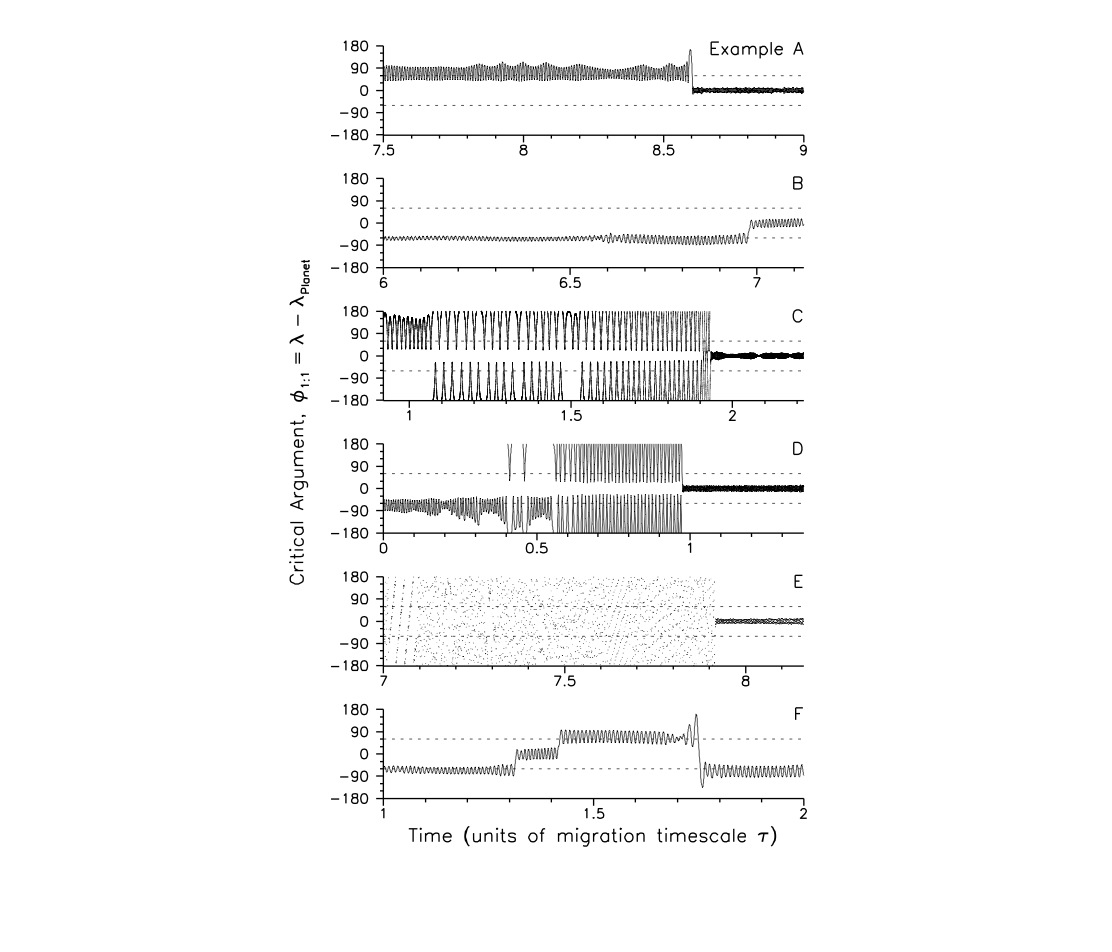}}}  
        \parbox[10cm]{16.5cm}{
        \renewcommand{\baselinestretch}{1.0}
        \mysize \sf \makebox[0cm]{}\\[-0cm]
        Figure~\ref{FIG:qs_trapping_exp}: 
	\CAPTIONqstrappingexp
        }     
\end{minipage} 
\clearpage

\noindent
\begin{minipage}[t][15.0cm][t]{17cm}    \refstepcounter{figs} \label{FIG:1_4_planets_nep}
        \sloppy 
        \vspace{-4cm}
        \centerline{  
          \hspace{-1cm}
          \scalebox{0.95}{ \includegraphics*[bb=0 0 550 550]{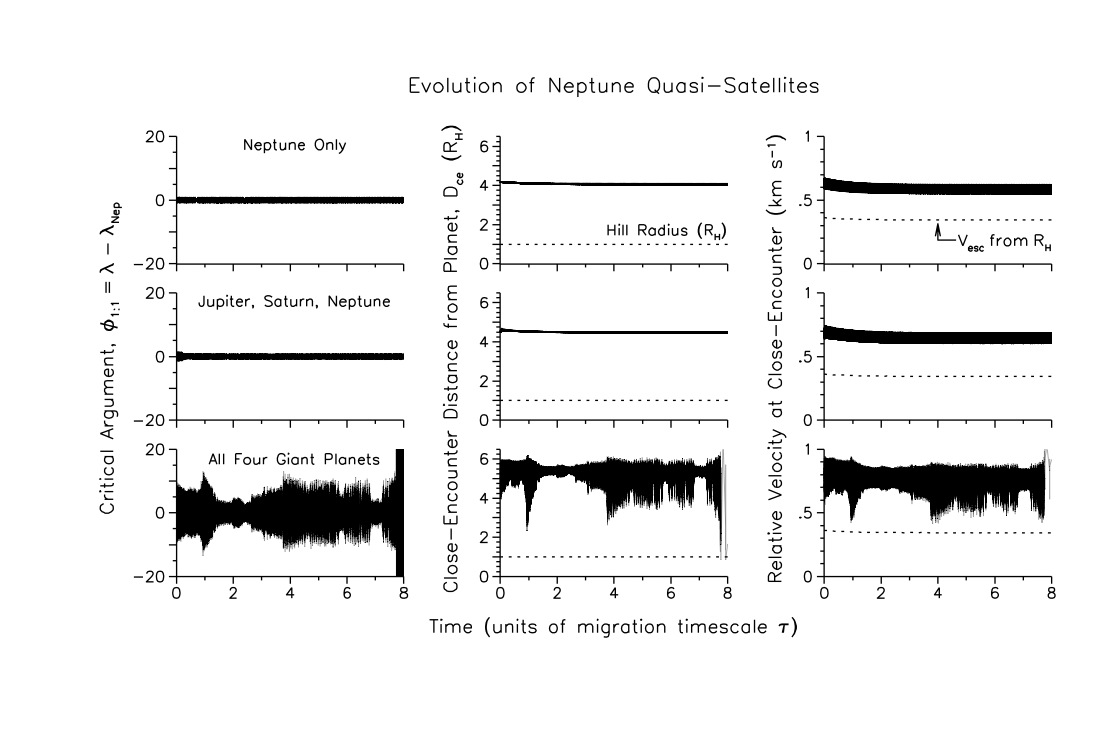}}}  
        \parbox[10cm]{16.5cm}{
        \renewcommand{\baselinestretch}{1.0}
        \mysize \sf \makebox[0cm]{}\\[-0cm]
        Figure~\ref{FIG:1_4_planets_nep}: 
	\CAPTIONonefourplanetexp
        }     
\end{minipage}
\clearpage

\noindent
\begin{minipage}[t][15.0cm][t]{17cm}     \refstepcounter{figs} \label{FIG:5tau_nep}
        \sloppy 
        \vspace{-3cm}
        \centerline{  
          \hspace{-0cm}
          \scalebox{0.95}{ \includegraphics*[bb=0 0 550 575]{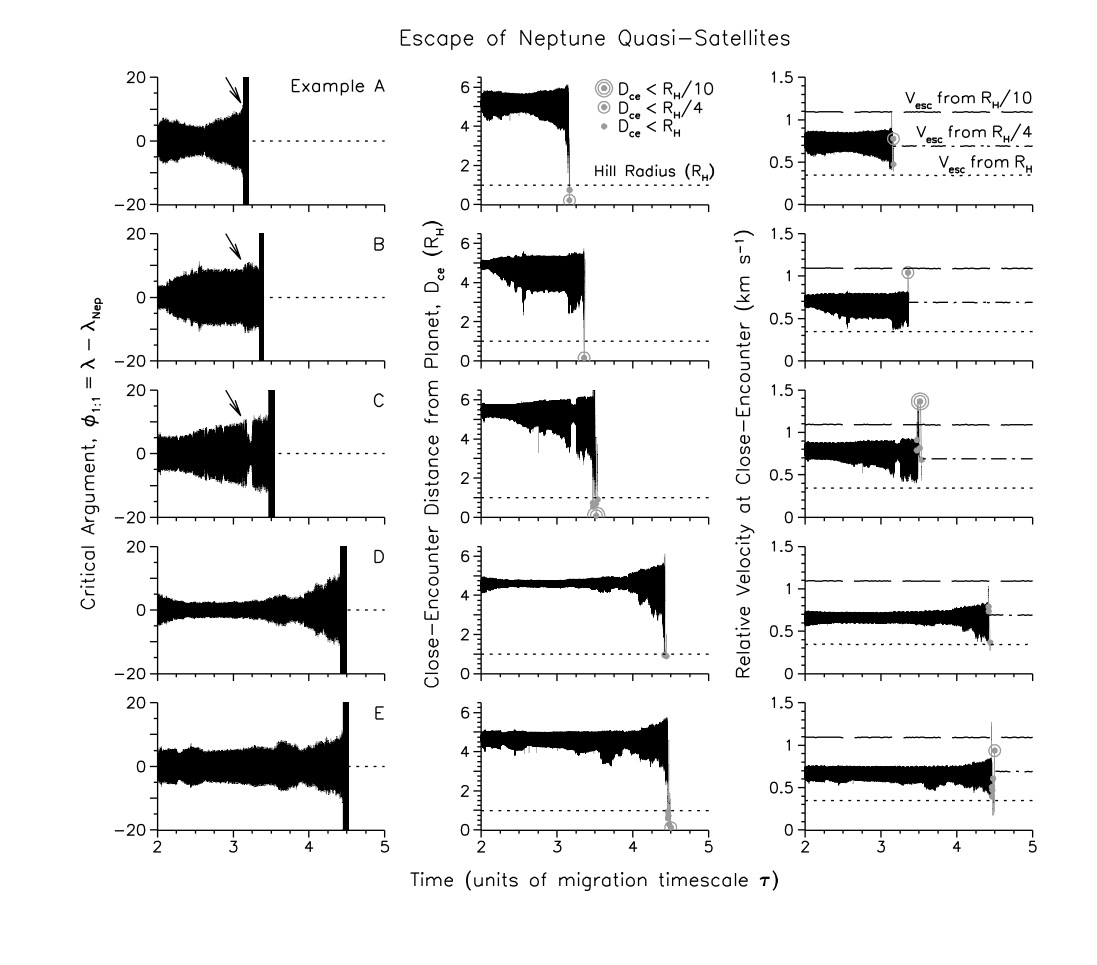}}}  
        \parbox[10cm]{16.5cm}{
        \renewcommand{\baselinestretch}{1.0}
        \mysize \sf \makebox[0cm]{}\\[-1cm]
        Figure~\ref{FIG:5tau_nep}: 
	\CAPTIONfivetaunep
        }    
\end{minipage} 

\noindent
\begin{minipage}[t][15.0cm][t]{17cm}    \refstepcounter{figs} \label{FIG:7tau_jup}
        \sloppy 
        \vspace{-3cm}
        \centerline{  
          \hspace{-0cm}
          \scalebox{0.95}{ \includegraphics*[bb=0 0 550 575]{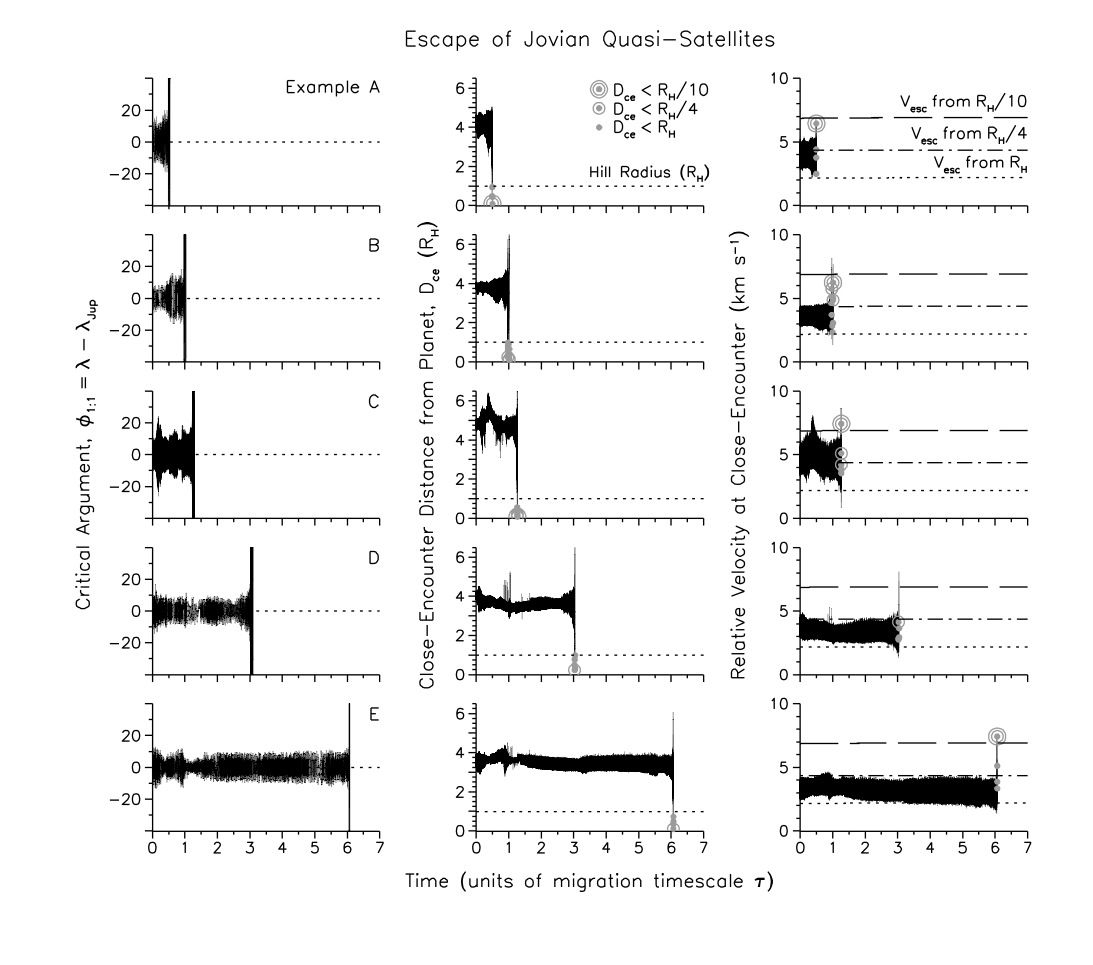}}}  
        \parbox[10cm]{16.5cm}{
        \renewcommand{\baselinestretch}{1.0}
        \mysize \sf \makebox[0cm]{}\\[-1cm]
        Figure~\ref{FIG:7tau_jup}: 
	\CAPTIONseventaujup
        }    
\end{minipage} 
\clearpage

\noindent
\begin{minipage}[t][15.0cm][t]{17cm}    \refstepcounter{figs} \label{FIG:L4_to_L5}
        \sloppy 
        \vspace{-0cm}
        \centerline{  
          \hspace{-1cm}
          \scalebox{0.8}{ \includegraphics*[bb=0 0 550 575]{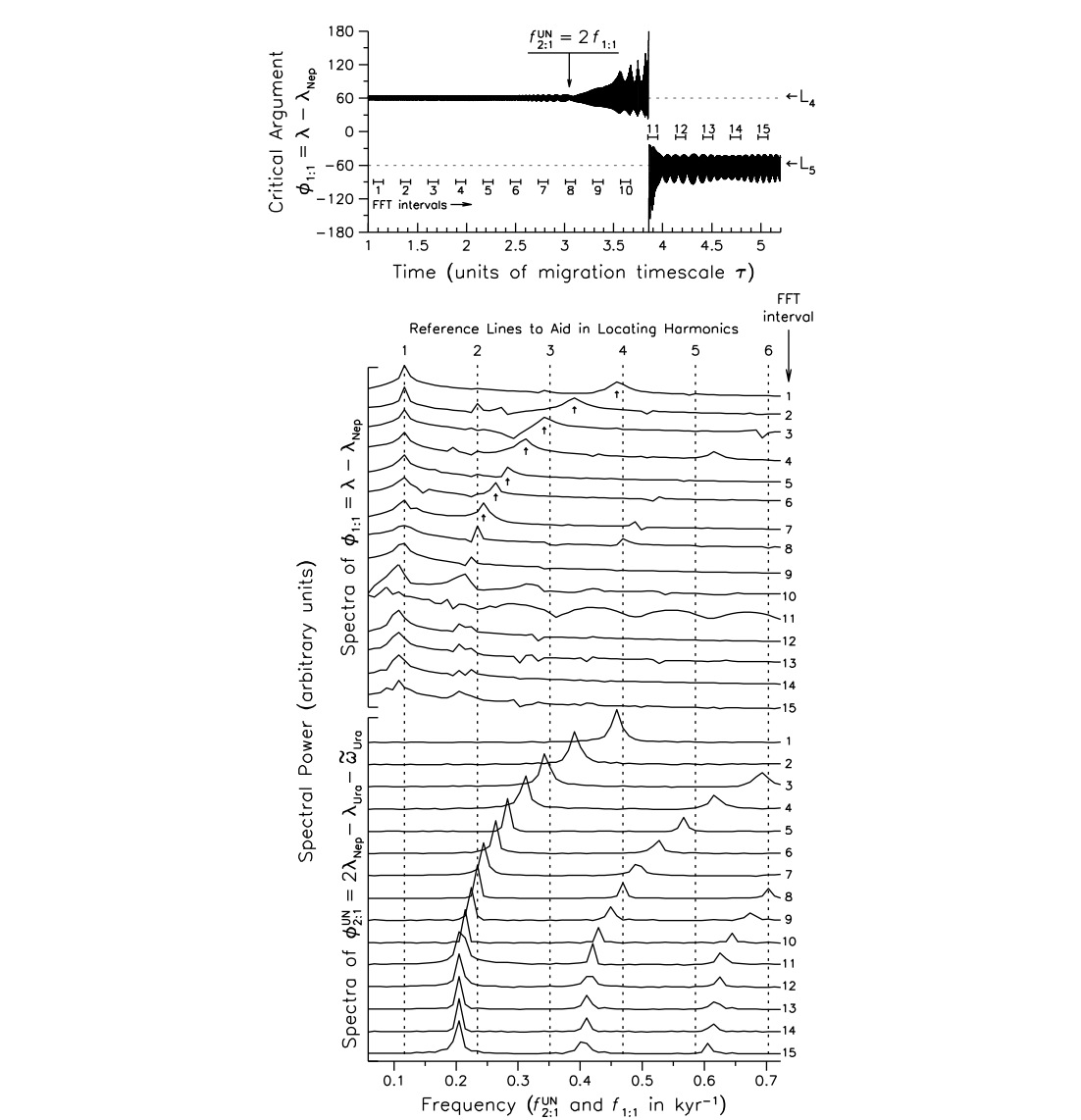}}}  
        \parbox[10cm]{16.5cm}{
        \renewcommand{\baselinestretch}{1.0}
        \mysize \sf \makebox[0cm]{}\\[1cm]
        Figure~\ref{FIG:L4_to_L5}: 
	\CAPTIONfourtofive
        }     
	\mylabel{last_total_page}
\end{minipage} 
\clearpage


\end{document}